\documentclass[jou]{apa}%can be [jou] (for journal), man (manuscript) or doc (document)  (was doc)  (article was apa)
\usepackage{url} %this allows us to cite URLs in the text
\usepackage{graphicx}  %allows for graphic to float when doing jou or doc style
\usepackage{amsmath,amssymb}
\usepackage{amsmath,amsthm}
\usepackage{subfigure}
\usepackage{rotating} 
\title{Automatic selection of eye tracking variables in visual categorization in adults and infants}
\author{Samuel Rivera$^{13}$, Catherine A. Best$^{23}$, Hyungwook Yim$^{23}$, Dirk B. Walther$^{4}$, Vladimir M. Sloutsky$^{23}$, Aleix M. Martinez$^{13}$}
\affiliation{$^1$Department of Electrical and Computer Engineering, The Ohio State University \\ 
$^2$Department of Psychology, The Ohio State University \\
$^3$Center for Cognitive Science, The Ohio State University \\
$^4$Department of Psychology, University of Toronto
}

\abstract{ Visual categorization and learning of visual categories exhibit early onset, however the underlying mechanisms of early categorization are not well understood. The main limiting factor for examining these mechanisms is the limited duration of infant cooperation (10-15 minutes), which leaves little room for multiple test trials. With its tight link to visual attention, eye tracking is a promising method for getting access to the mechanisms of category learning. But how should researchers decide which aspects of the rich eye tracking data to focus on? To date, eye tracking variables are generally handpicked, which may lead to biases in the eye tracking data. Here, we propose an automated method for selecting eye tracking variables based on analyses of their usefulness to discriminate learners from non-learners of visual categories. We presented infants and adults with a category learning task and tracked their eye movements. We then extracted an over-complete set of eye tracking variables encompassing durations, probabilities, latencies, and the order of fixations and saccadic eye movements. We compared three statistical techniques for identifying those variables among this large set that are useful for discriminating learners form non-learners: ANOVA ranking, Bayes ranking, and L1 regularized logistic regression. We found remarkable agreement between these methods in identifying a small set of discriminant variables. Moreover, the top eye tracking variables allow us to identify category learners among adults and 6- to 8-month-old infants. \\

\noindent keywords: eye tracking, infant category learning, eye tracking variables
}
\begin{document}
\maketitle

\section{Introduction}
Categorization is the process of forming an equivalence class, such that discriminable entities elicit a common representation and/or a common response.  While category learning exhibits early onset \cite{Quinn1993}, relatively little is known about the underlying mechanism and the development of early categorization.  The primary reason is the limited duration of infants' cooperation, yielding only a small number of data points per participant. These limitations have restricted researchers' ability to answer fundamental questions about categorization in infants:  How do infants learn a category? And does this process undergo development?  

Analyses of eye movements may help solve some of these problems: eye movements are tightly linked to visual attention (see Rayner, 1998, for a review) and they yield multiple (albeit not necessarily independent) data points even for relatively short trial durations.  Therefore, analyses of eye movements can provide critical information of how attention allocation changes in the course of category learning.  However, eye tracking results in a large amount of data, and it is not clear a priori what (if any) components of eye movement are related to category learning.  As a result, eye tracking researchers are free to choose from a large set of variables without a common set of principles for deciding which or how many variables to analyze.  In the following, we review the infant category learning eye tracking literature in order to substantiate this claim.  Given the limited number of eye tracking categorization studies with infants, we take a broader approach and review studies that examined categorization, object completion, and visual attention. 

One variable that has been used across a variety of tasks is saccade latency \cite{amso2006,johnson2003}.  For example, Johnson, Amso, and Slemmer (2003) examined whether learning affects object representations in infancy. Four- and six-month-old infants were presented with an object that moved behind an occluder and then reemerging on the other side of the occluder.  The researchers reasoned that if babies maintain the existence of the occluded object, they should anticipate the object to reemerge from the occluder.  In this case, participants should exhibit a faster eye movement to the point of reemergence than if they do not anticipate the object.  In two other studies \cite{amso2005,amso2008}, researchers used saccade latency to examine the development of visual selection.  Participants (3-, 6-, and 9-month-olds and adults) were presented with a Spatial Negative Priming (SNP) paradigm.  On a given trial they were shown an attention grabbing target in Location 1 and a discreet distracter in Location 2.  On the next trial, they were either shown the target in Location 2 (a negative priming probe) or in Location 3 (a control trial).  SNP was inferred from greater saccade latency on the probe trials than on the control trials.

Other potentially informative variables are (a) frequencies of fixations per unit of time within one or more Areas of Interest (AOIs), (b) dwell times within one or more AOIs, and (c) frequencies of saccades within and between AOIs \cite{johnson2004}.  In one study, Johnson and colleagues (2004) examined the development of object unity perception in infancy using both behavioral and eye tracking data. In the task, participants were habituated to a rod moving behind an occluder. After participants habituated, the occluder was removed to reveal either a broken rod or a complete rod.  Infants who perceived the rod as moving behind the occluder as a coherent object, indicated by a recovery of looking after habituation, were identified as perceivers. Participants who perceived a broken rod did not recover their looking after habituation and were considered non-perceivers.  The authors then examined eye tracking data for perceivers and non-perceivers using the eye tracking variables described above.

Perhaps the most frequently used eye tracking variable is fixation location. Researchers have relied on this variable across a variety of tasks, including object completion \cite{amso2006,johnson2008}, understanding other people's actions \cite{Falk-Ytter:2006}, and a variety of categorization and category learning tasks \cite{best2010,McMurray:2004,Quinn:2009}.  For example, Quinn et al. (2009) examined categorization of cats and dogs in 6- to 7-month-olds.  They found that when items were presented in the canonical upright position, categorization accuracy was associated with a high proportion of looking to the head; whereas, when items were presented in an inverted position, categorization was associated with the large proportion of looking to the body.  Best et al (2010) presented 16- to 24-month-olds with a category learning task.  Categories included artificial items that had shapes in four locations, with two of the shapes being category relevant (i.e., present in all members of the category, but not in non-members) and two being irrelevant (i.e., exhibiting both within- and between-category variability).  The researchers examined the proportion of fixations to category-relevant features and its change in the course of familiarization.  A summary of the reviewed studies is presented in Table \ref{table:chart}.  

Several conclusions can be drawn from this brief review.  First, multiple eye tracking variables have been used across studies to examine infants' learning.  And second, although all these variables make intuitive sense, no formal selection process of these variables has been defined.  This poses several concerns and questions.  Namely, since different variables are used in different studies, do these variables correlate and thus provide redundant information?  If not, why should any one variable be used instead of another?  Should the variables be selected based on the specific categorization task, or should a fixed subset of eye tracking variables be used across all studies?  Can we define a principled way of determining which variables to analyze in a given category learning study?  The current study defines a methodology to address these questions and concerns. 
   
\begin{table*}
%\centering
\scriptsize
%\small
\begin{tabular}{  l  l l }
\hline
{\bf Source} & {\bf Task } & {\bf Eye Tracking Variable }\\  \hline
Johnson, et al., PNAS, 2003 & Object completion & Saccade latency \\ 
Amso and Johnson, Developmental Psychology, 2006 & Object completion & Proportion fixation to AOI \\ 
Johnson, et al, Infancy, 2004 & Object completion & Fixation frequency, dwell time, saccade frequency \\ 
Johnson, et al, Developmental Psychology, 2008 & Object completion & Proportion fixation to AOI \\ 
Falck-Ytter, et al, Nature Neuroscience 2006 & Goal perception & Proportion fixation to AOI, AOI fixation time \\ 
Amso and Johnson, Cognition, 2005 & Visual Search & Saccade latency \\ 
Amso and Johnson, Infancy, 2008 & Visual Search & Saccade latency \\ 
Quinn et al, Child Development, 2009 & Categorization & Proportion fixation to AOI \\ 
McMurray and Aslin, Infancy, 2004 & Category Learning & Proportion fixation to AOI \\ 
Best, Robinson, and Sloutsky, Proceedings of Cognitive Science Society, 2010 & Category learning & Proportion fixation to AOI \\  \hline
\end{tabular}
\caption{Comparison of previous eye tracking variables.  }
\label{table:chart}
\end{table*}

Our approach was as follows. We extracted a large set of possible variables from the adult or infant gaze sequence during a categorization task (e.g. fixations, saccades, gaze sequences, etc).  Some of the variables have been used in analyzing categorization experiments, whereas others were new. Our goal was to use the power of statistics and machine learning to identify eye tracking variables that best predict category learning in adults and subsequently in infants.  The significant contribution of this work is that it provides a systematic methodology for identifying eye tracking variables that are linked to category learning, thus allowing researchers to better understand category learning from eye tracking data.  Futhermore, our results retrospectively validate the use of several variables from the eye tracking studies mentioned above.  
 
\section{Methods}  
\subsection{Participants}  
Three category learning experiments were conducted where two focused on adults and one on infants.  Twenty-four adults participated in Experiment 1.  Forty-six adults who did not participate in Experiment 1 participated in Experiment 2.  All adult participants had normal or corrected to normal vision and were undergraduate students at The Ohio State University participating for course credit.  

In Experiment 3, fifteen 6- to 8-month-old infants participated in the experiment. Parents provided written consent upon arrival to the laboratory. All parents reported their infants to be developing typically and in good health. 

\subsection{Materials} 
\begin{figure}
\centering
\subfigure[Category object]{
\includegraphics[height=2cm]{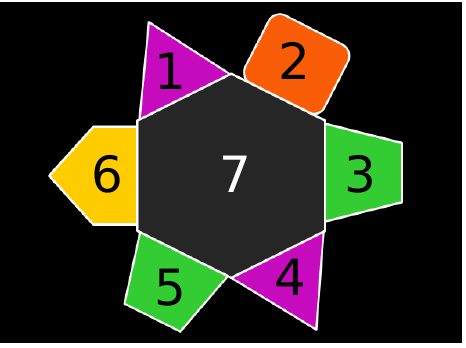}
\label{Fig:categoryExample}
}
% \subfigure[AOI Labels]{
% \includegraphics[height=2cm]{images/exampleCat2_labeled.pdf}
% \label{fig:stick}
% }
%\hspace{.5cm}
\subfigure[AOI Example ]{
\includegraphics[height=2cm,width=2.2cm]{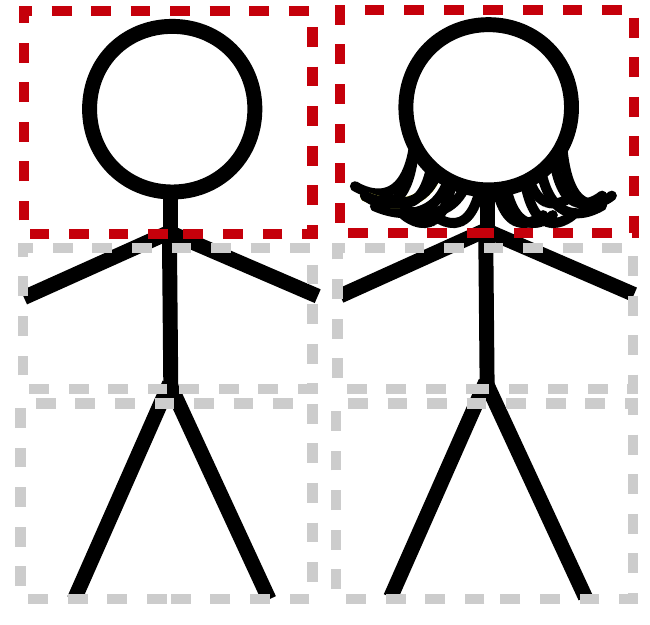} 
\label{fig:stick}
}
%\subfigure[Category object pair ]{
%\includegraphics[height=2cm]{images/exampleCat2_labeledPair.pdf} 
%\label{fig:pair}
%}
% \vspace{-0.5cm}
\caption{\small Image \subref{Fig:categoryExample} is an example category object used in the eye tracking study, with the Areas of interest (AOI)s enumerated. Numbers were not displayed to the participants. Image \subref{fig:stick} illustrates the concept of AOIs.  Each stick figure is divided into $3$ AOIs containing the head, torso, and legs.  The relevant AOI for \emph{gender} discrimination is bracketed in red.  Only the head AOI is relevant because the other AOIs are the same across both categories. }
\label{Fig:CatAndStick}
\end{figure}

Category members were flower-like objects with six petals. An example object is shown in Fig. \ref{Fig:categoryExample}, with the petals enumerated for clarity. There were four different categories, each defined by a single petal having a distinguishing color and shape. Specifically, the category defining features were category A: a pink triangle at position 4; category B: a blue semi-circle at position 4; category C: an orange square at position 6; and category D: a yellow pentagon at position 6. Each object was uniquely associated with one category. That is, no one object exhibited the defining features for two or more categories.  Stimuli were displayed on the computer subtending an approximate horizontal visual angle of 11$^\circ$ and an approximate vertical visual angle of 11$^\circ$. The eccentricity of the stimuli subtended an approximate horizontal visual angle of 14.4$^\circ$ and an approximate vertical visual angle of 11.5$^\circ$. 

During all three experiments, the participants' eye gaze was recorded using a Tobii T60 eye-tracker (Falls Church, VA) at the sampling rate of $60$Hz while they sat approximately $60$ cm away from the display screen.  

\subsection{Experiment 1 - Adult supervised}\label{sec:adultExp} 
To validate the efficacy of the approach before applying it to infants, adult participants were tested.  In Experiment 1, participants were instructed to look for a single distinguishing feature prior to the start of the experiment. Previous research suggests that this hint (i.e., a form of supervised learning) has large consequences with respect to how quickly participants learn to classify the objects, especially when there are few overlapping features \cite{Kloos2008}. 

The experiment had $8$ blocks where in each block there were $8$ learning trials followed by $4$ testing trials. In a learning trial, a category member was displayed in the center of the screen one at a time for $1.5$ seconds each. In a testing trial, a novel category member of the to-be-learned category and a novel member of a contrasting category were presented side by side each centered approximately 7.7$^\circ$ horizontal visual angles from the center of the screen. Test stimuli were displayed on the screen until the participant made a decision via key press about which stimulus was a member of the learned category. The left/right position of the test stimuli  was counter-balanced.  A randomly located fixation point (cross-hair) directed the participant's gaze to a position on the monitor in-between trials. The to-be-learned category remained the same for the first $4$ blocks. A second to-be-learned category was introduced in the final $4$ blocks without notice to the participant. If the experiment started with a category defined by the petal at position 4 (category A or B), the second category was defined by the petal at position 6 (category C or D), and vice-versa. Using categories having definitive features at different positions provided a mechanism to verify the reproducibility of the variables determined most important.  

\subsection{Experiment 2 - Adult unsupervised}\label{sec:adultExp} 
The procedure in Experiment 2 (unsupervised condition) was identical to that in Experiment 1 except that participants did not receive supervision (i.e., no hint provided) about the category structure.
\begin{figure}
\centering
\includegraphics[height=2cm]{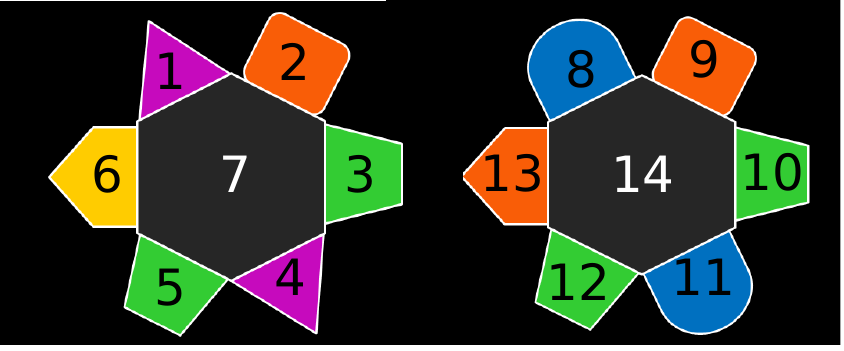} 
\caption{ \small Illustration of category pair image with AOIs labeled.  Numbers were not shown to participants. }
\label{Fig:pair}
\end{figure}

\subsection{Experiment 3 - Infant supervised}\label{sec:infantExp} 
 The infant experiment was conceptually similar to Experiment 1, but was methodologically adapted for infants by using a familiarization paradigm.  To aid infant learning, category exemplars were shown in pairs on each trial.  This was also done so that the presentation of stimuli in the learning and testing phases had an identical layout.   An example with labeled AOIs is shown in Fig. \ref{Fig:pair}. Furthermore, there was only a supervised condition, in which the infants were presented with a pre-trial fixation video of synchronized sound and motion (e.g., looming flower petal with corresponding whistle sound) to draw their attention to the single category-relevant feature.  It should be noted that no unsupervised condition was conducted with infants because previous developmental research suggests supervision is necessary for young children to learn categories with a sparse category structure \cite{Kloos2008}.  Once the infant looked at the fixation video, the learning trial commenced.  Infants had to accumulate 3 seconds of looking to the category exemplar pairs.  Whenever an infant looked away, an attention-grabbing fixation was presented until the infant reconnected with the images on the screen.  After accumulating 3 seconds of looking to the stimulus pair, the supervisory fixation video was again presented followed by another learning image pair.  This procedure was repeated for $8$ blocks with $8$ learning pairs per block.

In the testing phase, a novel category member was paired with a novel non-category member as in the adult experiments.  The standard assumption is that an infant can discriminate between the category and non-category objects if he or she displays a novelty or familiarity preference.  There were two test trials per block, where a novel exemplar from the learned category was paired with a novel exemplar from a novel category. Test trials were presented for a fixed duration of 6 seconds, and left/right position of familiar or novel category objects was counterbalanced.    

\subsection{Collecting and filtering eye tracking data} 
Eye movements were monitored during object viewing with the Tobii T60 eye tracker. The system tracks eye movements by illuminating the eye with infrared light and capturing corneal reflection at a frequency of 60 Hz (i.e., every 16.6 ms).  As the eye moves, the angle between the pupil and the corneal reflection increases, allowing the x-y coordinates of the gaze position to be measured over time.  

Unfortunately, the gaze data contain noise, missing data, and micro-saccades, which makes identifying true fixations and saccades difficult.  Therefore, we processed these data using MATLAB-based software created in our laboratory by the first author.  The raw eye tracking data from every experimental block were filtered using a Kalman filter \cite{kalmanToolbox} before extracting the variables of interest.  The eye gaze data from both the left and right eye were filtered separately. The average of the filtered data from left and right eyes yielded the mean eye gaze data, which were used in the current analyses.    

\subsection{Labeling the Data}\label{section:labelData}
The eye movement sequences during the \emph{learning phase} of the experiment aid in understanding category learning, while the sequences during the \emph{testing phase} aid our understanding of category use.  Before applying our methodology to understand these processes, however, the eye tracking data from both the learning and testing phases of the experiments were labeled as \emph{learner} (class 1), \emph{non-learner} (class 0), or \emph{indeterminate} (class 2).  Indeterminate samples were not analyzed. 

\begin{figure}
\centering
\includegraphics[width=0.25\textwidth]{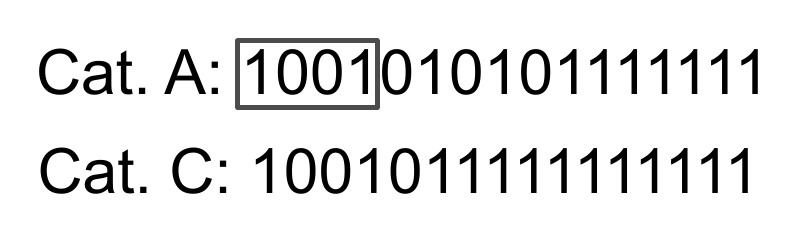}
% \vspace{-.3cm}
\caption{ \small Illustration of a time series for one subject.  Ones encode correct category discrimination, while zeros encode incorrect responses.  The first row shows the accuracy over the first four blocks (presentation of first category), while the second row shows accuracy over the last four blocks (presentation of second category).  The class labels (learner or non-learner) are determined separately for each row, because the category condition is different for each row. }
\label{fig:respString}
\end{figure}

\noindent{\bf Adult Labels}: 
Intuitively, labels for adult data are readily identified based on the accuracy of the responses during the testing phase.  An uninterrupted string of correct responses during the testing phase suggests that the participant has learned the category.  
Each adult experimental block yielded $12$ eye movement sequences.  These correspond to eye movements during the presentation of $8$ exemplar images during the learning phase and $4$ test images during the testing phase.  Adult participants had $4$ blocks of learning and discriminating the same category before switching to a new category.  This amounted to $32$ samples of the learning phase, and $16$ samples of the testing phase for each category per participant.  The $16$ samples from the testing phase were associated with a $16$ digit binary string, called the \emph{response string}.  This data structure shows performance over the first and last $4$ blocks of the experiment.  A one identifies a correct response, while a zero denotes an incorrect response on the associated test trial.  An example is shown in Fig. \ref{fig:respString}.  We labeled each $16$ digit response string separately as follows.

We expect a learner's response string to contain a series of ones beginning within the string and terminating at the end of the response string.  This pattern indicates that at some point the participant learned the category and correctly discriminated the category from that point on.  A participant who has not learned the category (non-learner) would select one of the two stimuli by chance in each trial. A non-learner could get lucky and achieve a series of correct guesses. In order to determine if a participant is a learner or a non-learner we need to establish a criterion that allows us to reject chance as the cause for a series of ones. The question that we need to answer is how many ones we should expect for a learner.  We address this problem  by assessing how likely it is that we see a sequence of $M$ consecutive ones in a binary response string of length $R = 16$.  Under the null hypothesis, the participant does not know the category label and selects one of the stimuli by chance, giving her a $50\%$ chance of correctly guessing the category member.  Each sequence is equally likely given this assumption, so the probability of guessing at least $M$ right in a row is the total number of sequences having $M$ ones in a row (  $(R-M+1)\times2^{(R-M)}$ ) divided by the total number of binary sequences of length $R$ ( $2^R$).  This yields the probability $p= (R-M+1)/(2^M)$.  For $R=16$, $M = 10$ is the minimum number that achieves a significance level of $p < 0.01$ ($p =0.0068$ ).   Therefore, we rejected the null hypothesis that a participant was guessing randomly when we identified a consecutive string of $10$ correct responses.   

We call the position of the first correct response in this string of correct responses the point of learning (POL).  The \emph{test phase} and \emph{learning phase} samples before the POL were labeled as non-learner, while the samples after the POL were labeled as learner.  The learning phase samples from the block associated with the POL were labeled as indeterminate, because it was unclear at exactly which trial during the block the category was learned.  

If the learning criterion was not achieved, we then identified the remaining non-learning and indeterminate samples.  We first labeled correct responses at the end of the respond string as indeterminate.  Those samples did not meet the learning criterion, but might be attributed to learning late in the experiment.  The remaining samples were labeled as non-learner.  Approximately $8\%$ of the adult eye track samples were labeled indeterminate.

\noindent{\bf Infant Labels}: Obviously, infants are not able to respond by keyboard to identify a category object.  Instead, we used a variant of the preferential looking paradigm to determine if an infant could discriminate between novel exemplars of a familiar category object and a novel category object.  Recall that the preferential looking paradigm assumes that infants who consistently look more to one class of stimuli when shown two classes of stimuli are able to discriminate between the two classes.  This means that if the infant consistently looks longer at the learned category object (or novel category object), then he or she is assumed to be discriminating between the familiar and novel categories.

Given this paradigm, we labeled each infant's gaze data by blocks.  Each block consisted of two test phase samples.  We determined novelty preference as the ratio of total looking time to the novel category object compared to the total looking time to the novel category plus the familiar category object.   We sorted the mean of the novelty preference for each block according to the absolute difference from $0.5$.  A third of the blocks with mean novelty preference closest to $0.5$ were labeled as non-learner.  The third of the blocks with novelty preference furthest from $0.5$ in absolute value were labeled learner.
 Otherwise, the samples were labeled indeterminate.  Approximately $33\%$ of the infant eye track samples were labeled indeterminate.

\subsection{Variable List}\label{variablesList} 
We compiled an over-complete list of eye tracking variables.  We began with the fundamental variables, fixations and saccades.  Fixations occur when eye gaze is maintained at a single position for at least $100$ms. They were identified using the dispersion threshold algorithm of \cite{Salvucci2000}.  Saccades are rapid eye movements that move the eye gaze between points of fixation.  To be considered a saccade, the eye movement needed to exceed smooth pursuit velocity of $30^{\circ}$ per second or $0.5^{\circ}$ per sample at $60$Hz \cite{stampe1993}.  The fixations and saccades were determined with respect to a specific AOI within an object.  AOIs are regions of an object image or scene that can be grouped in some meaningful way, such as color uniformity or the structural nature of the object.  The AOIs can further be described as relevant or non-relevant, based on their role in determining object category membership.  Fig. \ref{fig:stick} illustrates this concept for the stick-figure \emph{gender} category.  In this toy example only the head is relevant for category membership because the torso and legs are identical across stick-figures and thus do not help one to discriminate gender.  

\begin{figure}
\centering
\subfigure[Single object]{
\includegraphics[height=2.6cm]{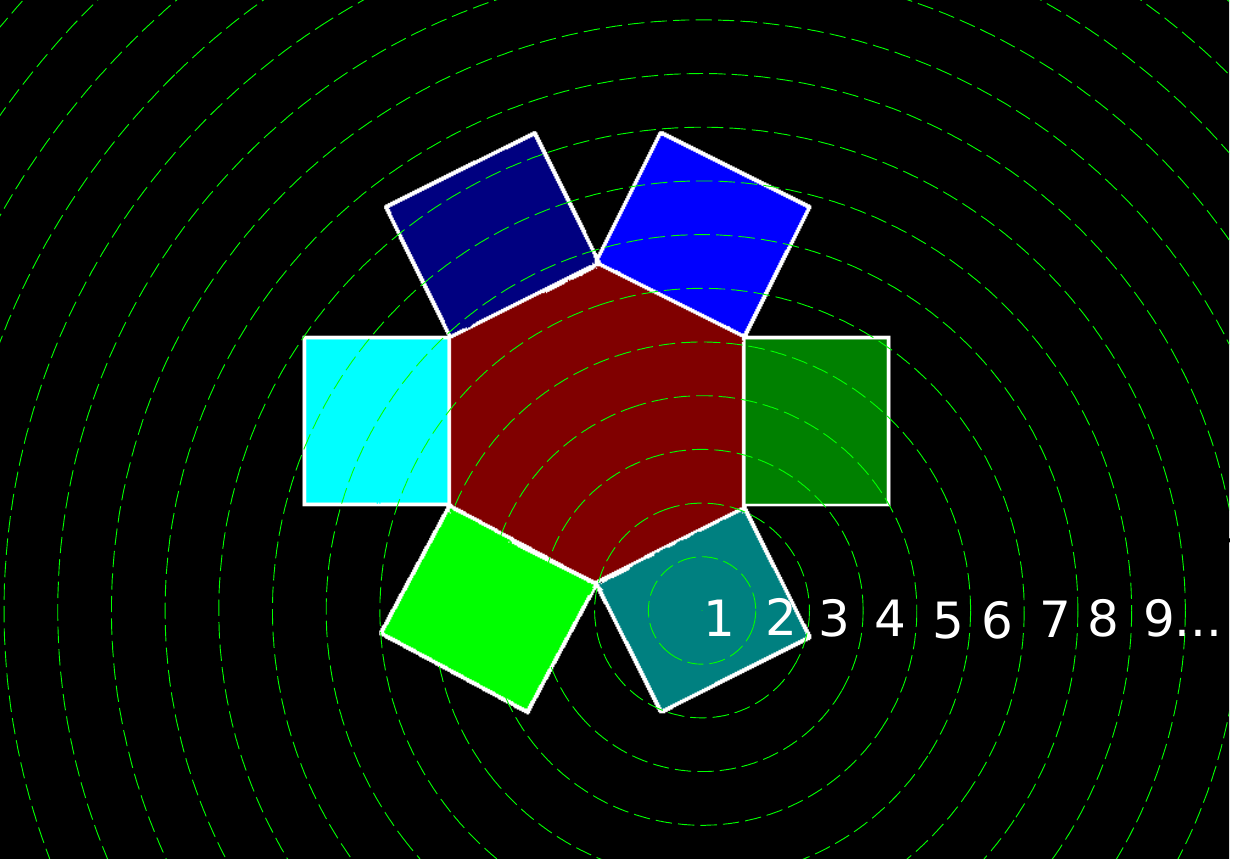}
\label{fig:learnHist}
}
\subfigure[Object pair ]{
\includegraphics[height=2.6cm]{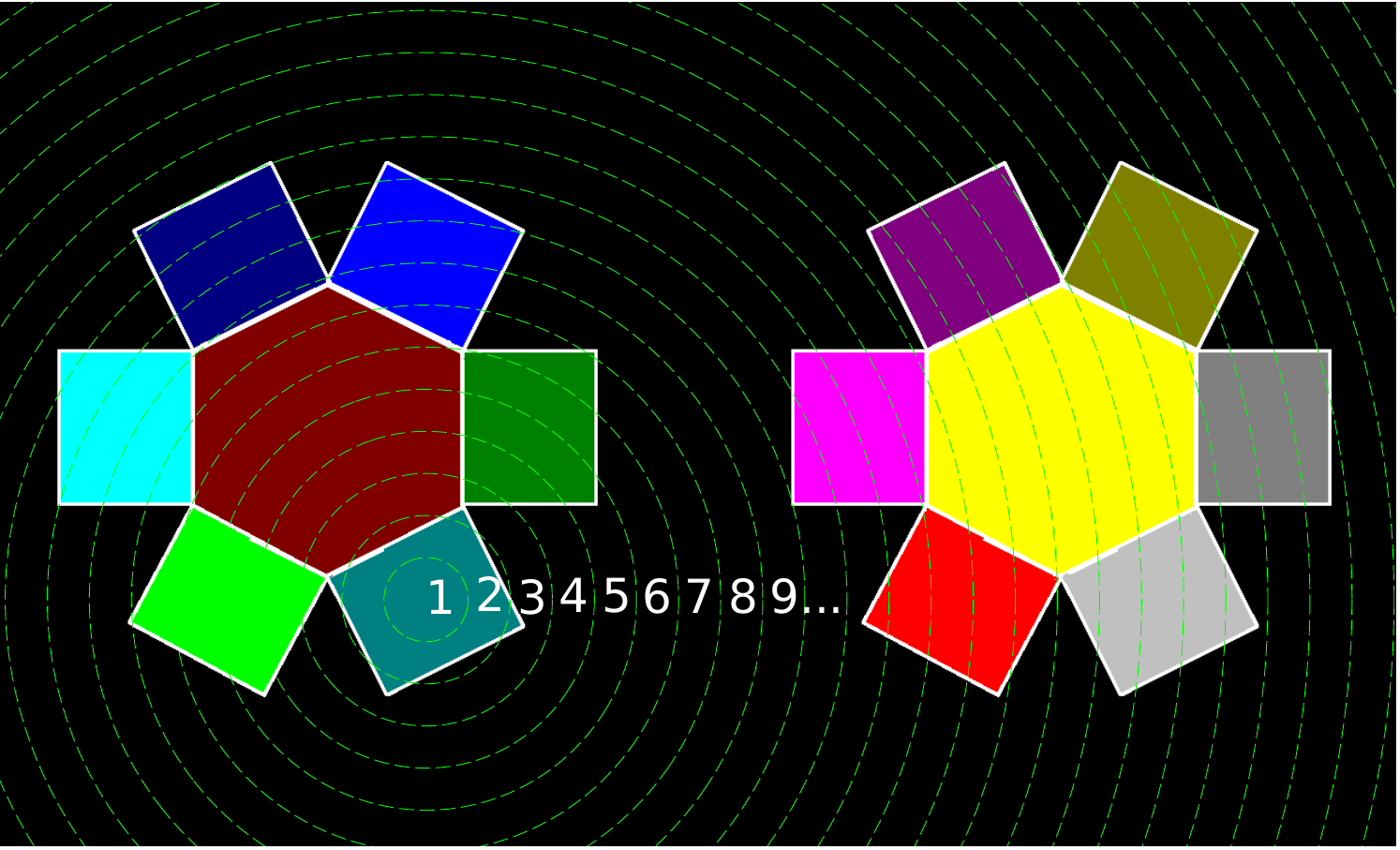} 
\label{fig:testHist}
}
\caption{\small Illustration of the histogram bins for distance to AOI 4, with bins numbered.  Variable $6$ describes the percent of time fixating within each bin for each relevant AOI.  Bin sizes were determined using CV. }
\label{Fig:histBin}
\end{figure}

These fundamental eye tracking variables were combined in various ways to derive a larger set of variables.  Our variable list is defined as follows:
\begin{enumerate}
   \item \emph{AOI fixation percentage} describes the percentage of time fixated at the different AOIs during a trial.  All non-AOI fixations were discarded in this and all of the variables defined. For an image with $q$ AOIs, this variable was encoded as a $q$-dimensional feature vector with a value for each AOI.  The fixation percentages were normalized so that they sum to $1$, unless there were no fixations at AOIs.  In that case, all percentages were set to $0$. 
   \item \emph{Relevant AOI fixation density} is a scalar value between zero and one which describes the percentage of the total time fixated which is at the relevant AOI(s). 
   \item \emph{AOI fixation sequence  } describes the sequence of AOI fixations during one trial. We limited this sequence to seven fixations, starting with trial onset (not counting fixations to the fixation mark).  We encoded a fixation sequence of $f$ fixations over $q$ AOIs as a $q \times f$ binary matrix, where each column of the matrix had a $1$ in the position corresponding to the AOI which was fixated, and zero otherwise.  If there were less than $f$ fixations, the last columns were set to $0$.  This binary encoding of the fixation sequence allowed us to describe any sequence of fixations without imposing an ordering of the AOIs.    In addition, the fixation sequence was represented as a sequence of relevant and non-relevant AOI fixations.  This representation yielded a $2 \times f$ binary matrix, in which each column had a $1$ in the first row if a relevant AOI was fixated or a $1$ in the second row if a non-relevant AOI was fixated. If there were less than $f$ fixations, the last columns were set to $0$.  The analysis showed that the latter representation was more informative in some cases.  Note that it was necessary to use a pair of binary variables to encode each fixation of the latter representation because it allowed for three cases: fixation at a relevant AOI or non-relevant AOI, and less than $f$ fixations.  The number of fixations to consider as well as the start position were determined using cross validation (CV).  In cross validation, the training data is separated into $k$ partitions, and for each partition, samples are classified using a classifier that is trained with the remaining $k-1$ partitions.  The percentage of correctly classified samples over all partitions is the CV accuracy.  We varied the parameters (start fixation and number of fixations) and calculated the CV accuracy when using only the fixation sequence to classify samples.  Using the first few fixations gave the best results, with no improvement as later fixations were included.  
    \item \emph{Duration of fixations in sequence} describes the duration of each fixation in the sequence described by variable $3$. This variable was encoded by an $f$-dimensional vector. 
   \item \emph{Total distance traveled by eye} is a scalar describing the total distance traveled by the eye gaze during a trial.
   \item \emph{Histogram of fixation distances to relevant AOI} describes how much time is spent fixated near or far from the relevant AOI(s).  A histogram with $h$ bins and an image with $r$ relevant AOIs yielded an $h \times r$ dimensional matrix. Each column corresponds to a different relevant AOI, and each row corresponds to a particular range of distances from that AOI.  The entries define the percentage of time fixated at the distance ranges, so each column sums to 1.  If no fixations occured, all values were set to 0.  The number of bins was determined using CV.   The bins corresponding to AOI 4 are illustrated in Fig. \ref{Fig:histBin}. 
   \item \emph{Number of unique AOIs visited} is a scalar describing the total number of unique AOIs fixated during a trial.  AOI revisits were not counted as new. 
   \item \emph{Saccade sequence} is similar to variable $3$ but describes the sequence of AOI saccades during one trial.  All saccades whose targets were not to AOIs were discarded in this and all of the variables defined. The sequence was limited to seven saccades, starting at the first saccade.  The number of saccades to consider as well as the start saccade were determined using CV. We encoded a saccade sequence of $s$ saccades over $q$ AOIs as a $q \times s$ binary matrix. Each column of the matrix had a $1$ in the position corresponding to the AOI which was the target of the saccade, and zero otherwise.  If there were fewer than $s$ saccades, the last column(s) were set to $0$.  In addition, the saccade sequence was represented as a sequence of saccades to relevant and non-relevant AOIs.  This representation yielded a $2 \times s$ binary matrix, with each column containing a $1$ in the first row if saccading to a relevant AOI or a $1$ in the second row if saccading to a non-relevant AOI. If there were fewer than $s$ saccades, the last column(s) were set to $0$. 
      \item \emph{Relative number of saccades to an AOI } is the saccade analogue of variable $1$ and describes the relative number of saccades to the AOIs during one eye movement.  An image with $q$ AOIs yielded a $q$-dimensional feature vector with each entry counting the number of saccade targets at the corresponding AOI.  The vector was normalized by the sum of all entries such that the entries added  to $1$ unless there were no saccades. In that case, all entries were set to $0$.
      \item \emph{Fixation latency to relevant AOI} describes the delay before fixating at a relevant AOI during an eye movement. It was encoded as a scalar between $0$ and $1$, with $0$ corresponding to fixating to a relevant AOI immediately and $1$ describing a sequence with no fixation on a relevant AOI.   The value was computed as the start time of the first relevant AOI fixation divided by the total eye track time.
   \item \emph{Saccade latency to relevant AOI } describes the delay before a saccade to a relevant AOI.  It was also encoded as a scalar between $0$ and $1$, defined by the end time of the first saccade to a relevant AOI divided by the total eye gaze time.
\end{enumerate}

Thus, eye movements were represented by a \emph{feature vector} $\mathbf{x} = (x_1, x_2, \dots, x_d)^T$ whose $d$ entries correspond to the variables described.  Each feature $x_i$ was normalized to zero mean and unit variance over the entire dataset.  In addition, each $\mathbf{x}$ was associated with a class label, $y \in \{0,1\}$.  For clarity, \emph{features} denote the  entries of the feature vector which encodes the eye tracking variables, while \emph{variables} correspond to the measures of the eye tracking enumerated above.  Therefore, $d$ is much larger than $11$, because encoding certain variables requires multiple feature values.  Note that $d$ was the same for all feature vectors corresponding to images having the same number of AOIs and relevant AOIs because a fixed number of fixations and saccades were analyzed.

In the case of a single category object having one relevant AOI, variable $2$ is identical to one of the values of variable $1$.  Therefore, after extracting all variables from the gaze data of all participants, we did a simple redundancy check to eliminate cases of identical valued features.  For features $x_i,x_j$ to be identical, they must mirror each other over \emph{all} feature vectors for a particular category condition and within either the learning or testing phase.  In addition, the information encoded by several of these features overlaps. This over-complete representation allows us to find the encoding that is best suited to describe the categorization task. To this end, we performed variable selection on this over-complete set.

 %----------------------------------------------------------------------
\subsection{Variable Selection}\label{Section:ranking} 
Our goal was to identify the subset of variables from the set defined above that can best separate the classes: category learners and non-learners.  This was achieved using ANOVA feature selection by ranking, Naive Bayes Ranking (NBR), and L1 logistic regression (L1-LR).  
\\

\noindent{\emph{ANOVA feature selection}} relies on a standard hypothesis test on each feature of $\mathbf{x}$.  Specifically, let $x_i$ denote the $i^{th}$ feature of $\mathbf{x}$.  Using a dataset of eye tracking feature vectors and the associated class labels, we performed a two tailed $t$-test of the null hypothesis, which states that samples of $x_i$ coming from classes $1$ and $0$ are independent random samples from normal distributions with equal means, $\mu_{i1}$ and $\mu_{i0}$, respectively.  The alternative says that the class means are different.  We calculated the test statistics and the corresponding $p$-value.  A low $p$-value means the null hypothesis is rejected with confidence.  Since the goal was to find the variables which best separate the classes, the feature with lowest $p$-value was ranked as best.  
The $p$-values were calculated for all features $x_i, i = 1\dots d$, and they were ranked from best to worst according to increasing $p$-values.  
\\

\noindent{\emph{Naive Bayes Ranking}} (NBR) assumes that if the labeled feature vectors can be accurately classified given a single feature, $x_i$, then that feature separates the two classes well.  In essence, the classification accuracy is a surrogate for the class separability achieved by the particular feature.  Therefore, the features are ranked from best to worst according to decreasing classification accuracy. 

The Bayes classifier assigns a sample, $\mathbf{x}$, to the class having the highest posterior distribution.  More formally, assume that the class-conditional density functions of a feature, given its class, $p(x_i|y)$, are modeled as normally distributed with mean and variance, $\mu_{iy}$ and $\sigma_{iy}$, respectively.  Then by applying the Bayes formula, the posterior probability of class $y$ is $P(y|x_i) = \frac{p(x_i|y)P(y)}{p(x_i)}$, where $P(y)$ is the prior of class $y$, and $p(x_i)$ is a scale factor which ensures that the probabilities sum to $1$. In this work, we have $P(y=1) = P(y=0) = 0.5$, corresponding to the assumption that \emph{a priori} a sample is equally likely to come from a learner as from a non-learner. The scale factor is the same for both classes, so it can be omitted in the classification rule. Finally, the predicted class label, $\hat{y}$ is given by:
\begin{equation}
\hat{y} = \arg \max_{j \in \{0,1\}} p(x_i|y=j)P(y=j).
\end{equation}
% More formally, classify $\mathbf{x}$ as class $y=1$ if $p(x_i|y=1)P(y=1) > p(x_i|y=0)P(y=0).$  Otherwise, classify $\mathbf{x}$ as class $y=0$.      
\\

\noindent{L1 \emph{Logistic Regression} (L1-LR)} is a linear classifier model, which returns a probability that a sample belongs to a particular class.  It accomplishes this by modeling the natural logarithm of the ratio, or odds, of two probabilities as a linear function of $\mathbf{x}$.  More formally,
\begin{equation}\label{Equation:LRModel}
ln \left( \frac{p(y=1|\mathbf{x})}{1 - p(y=1|\mathbf{x})} \right) = \mathbf{w}^T \mathbf{x}-b,
\end{equation}
where $ln$ denotes the natural logarithm.  The two class probabilities are then given by
\begin{align}
p(y=1|\mathbf{x}) = & \frac{1}{1 + \exp(-\mathbf{w}^T \mathbf{x}+b)}, \nonumber \\
p(y=0|\mathbf{x}) = & \frac{\exp(-\mathbf{w}^T \mathbf{x}+b)}{1 + \exp(-\mathbf{w}^T \mathbf{x}+b)}. \nonumber
\end{align}
The parameters, $\mathbf{w}$ and $b$, are estimated via Maximum Likelihood (ML) estimation.  A regularization term $\lambda$ is introduced to penalize large elements in $\mathbf{w}$.  Using an L1-norm regularizer yields a sparse model.  More formally, the regularized ML objective is, 
\begin{equation}
\hat{\mathbf{w}} = \arg \max_{\mathbf{w},b} \sum_{i = 1}^N log P( y_i| \mathbf{x}_i) - \lambda \| \mathbf{w} \|_1,
\end{equation}
where $(\mathbf{x}_i,y_i),i = 1, \dots, N$ are the full feature vectors and their associated labels, $\lambda$ is a user determined real valued positive regularization parameter, and $\| \cdot \|_1$ denotes the L1-norm.  Increasing the value of $\lambda$ will result in more elements of $\mathbf{w}$ being shrunk to zero, i.e., more sparse.  Variable selection is performed by increasing the value of $\lambda$ until a desired number of $\mathbf{w}$ elements are non-zero.  The elements of $\mathbf{x}$ corresponding to the non-zero elements of $\mathbf{w}$ are the top ranked variables.  These top ranked variables can then be sorted from best to worst by sorting the corresponding entries of $\mathbf{w}$ in order of descending absolute magnitude.  We use the L1-LR implementation of \cite{L1toolbox}.

Each method results in a ranking of the features, $x_i$, from best to worst.  If we vectorize the indices of the $t$ top ranked features as $\mathbf{k} = (k_1, k_2, \dots, k_t)^T$, then after feature selection $\mathbf{x} = (x_{k_1}, x_{k_2} \dots, x_{k_t})^T$. 
 
%----------------------------------------------------------------------
\subsection{Linear Classification}\label{Section:classification}
\begin{figure}
\centering
\includegraphics[width=0.3\textwidth]{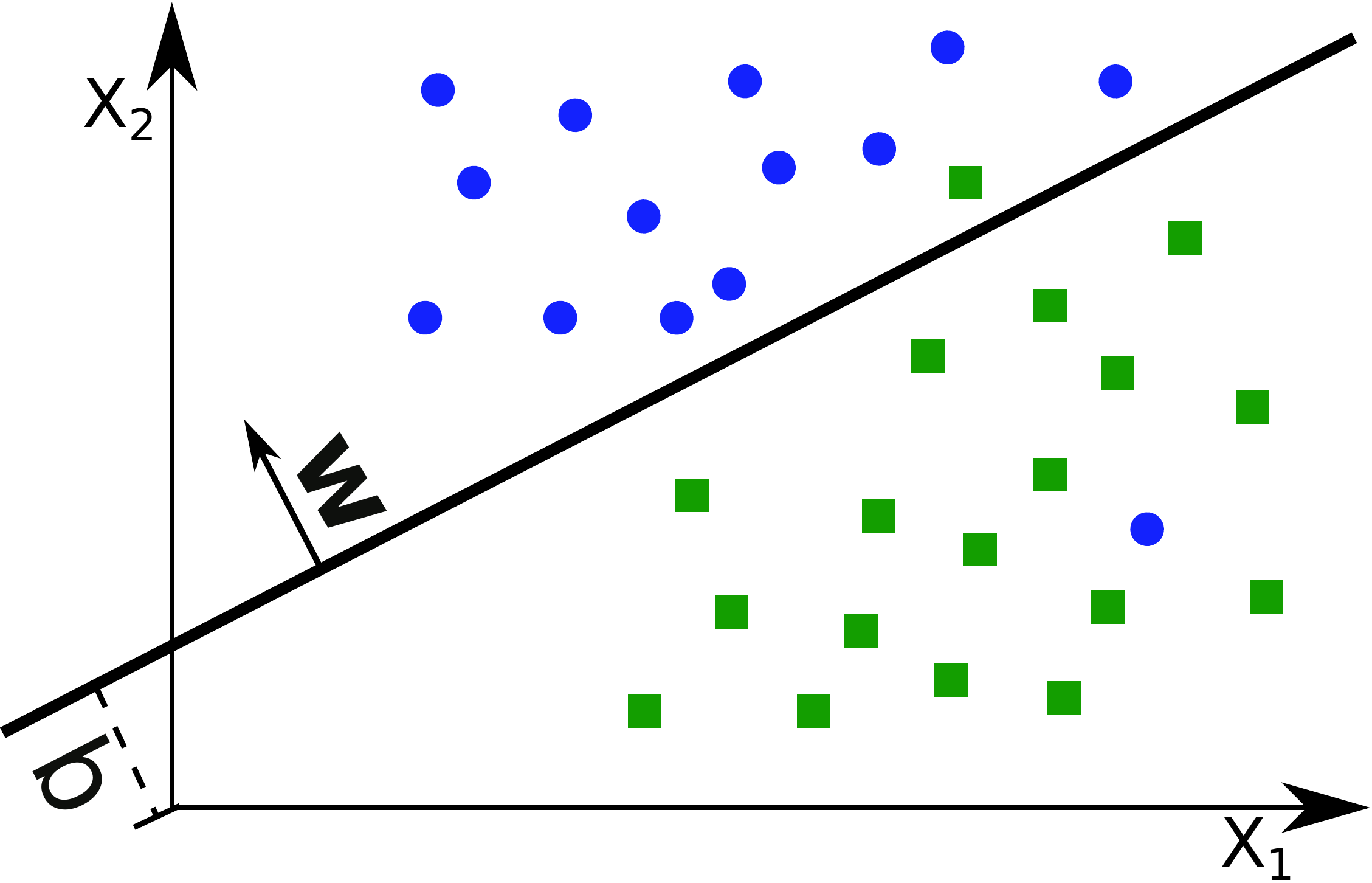}
% \vspace{-.3cm}
\caption{ \small Illustration of a linear classifier. $\mathbf{w}$ is the normal vector of the hyperplane which separates the feature space into two decision regions, and $b$ is the distance from the origin to the hyperplane.  The blue circles represent samples from class $1$, while the green squares represent samples from class $0$.  All but one of the blue circles exists on the positive side of hyperplane, and are classified correctly.   }
\label{fig:linClassifier}
\end{figure}

Once the important variables were identified, we used them to classify the gaze data as having originated from a learner or non-learner.  This required that we train a classifier to distinguish between two classes of data.  Recall that each eye movement resulted in a feature vector, or \emph{sample} $\mathbf{x}$. A classifier defines a decision rule for predicting whether a sample is from class $0$ or $1$.  A linear classifier was used because of its ease of interpretation \cite{martinez2005} -- the absolute model weights give the relative importance of the eye tracking variables.  We illustrate in Fig. \ref{fig:linClassifier} with a 2-dimensional linear classifier model specified by $\mathbf{w}$ and $b$.  $\mathbf{w}$ is the normal vector of the hyperplane which separates the feature space into two decision regions, and $b$ is the distance from the origin to the hyperplane (i.e., the offset). 

All samples $\mathbf{x}$ above the hyperplane are assigned to class $1$ while the samples below are assigned to class $0$.  Data samples $\mathbf{x}$ existing on the boundary satisfy $\mathbf{w}^T\mathbf{x}-b=0$.  Therefore, samples are classified according to the sign of $\mathbf{w}^T\mathbf{x}-b$.  In this example $\mathbf{w} = (-.55, .83)^T$, so the second dimension, $x_2$, is more informative for classification. Note that in our case the feature space has not two but up to $334$ dimensions, depending on the cut-off for variable selection.

Several varieties of linear classifiers exist.  In this work, we used the Bayes classifier with equal covariances, L1-LR, and the Support Vector Machine (SVM) algorithm.  \\

\noindent{\emph{Bayes with equal covariances (Bayes):}}  
When both classes are assumed to be multivariate normally distributed with the same covariance $\Sigma$, means $\mathbf{\mu}_1$ and $\mathbf{\mu}_0$, and equal priors, the Bayes classifier decision boundary is a hyperplane given by $\mathbf{w} = \Sigma^{-1}(\mathbf{\mu}_1 - \mathbf{\mu}_0)$ and $b = \frac{1}{2}\mathbf{w}^T(\mathbf{\mu}_1 - \mathbf{\mu}_0)$ \cite{duda2001}.   \\

\noindent{L1 \emph{Logistic regression:}} 
Recall that L1-LR yields a probability that a sample belongs to a particular class.  It uses the model of Equation \eqref{Equation:LRModel}, where $\mathbf{w}$ defines the normal of the hyperplane, and the sign of $\mathbf{w}^T\mathbf{x}-b$ determines the class label.\\

\noindent{\emph{Support Vector Machine:}} 
SVM is  a linear classifier which maximizes the margin between two classes of data \cite{Burges1998}.  In the case that the training samples are perfectly separable by a hyperplane, we can find $\mathbf{w}$ and $b$ such that the data satisfies the following constraints,
\begin{align}
\mathbf{x}_i^T \mathbf{w}-b & \geq 1 \text{~for~} y_i = 1,\\
\mathbf{x}_i^T \mathbf{w}-b & \leq -1  \text{~for~} y_i = 0.
\end{align}
Essentially, these constraints specify that the samples from the different classes reside on opposite sides of the decision boundary.  The margin between the classes, defined by $\frac{2}{\|\mathbf{w}\|_2}$ where $\|\cdot\|_2$ defines the L2-norm, is then maximized subject to the above constraints.  The dual formulation of the constrained optimization problem results in a quadratic program for $\mathbf{w}$ and $b$.  In the case that samples from each class are not linearly separable, a penalty is introduced to penalize the amount that a sample is on the wrong side of the hyperplane.  Again, the dual formulation results in a quadratic program for $\mathbf{w}$ and $b$.  We used the implementation of \cite{LIBSVM}.  

\subsection{Classification Accuracy}\label{Section:classification accuracy}
The classification accuracy used for adults was the leave-one-subject-out cross-validation (LOSO-CV) accuracy.  In LOSO-CV, the samples belonging to one participant are sequestered, and the remaining samples are used to train the classifier.  The sequestered samples are then classified with the learned classifier, and the procedure is repeated for every participant in the database.  The total number of correctly classified samples divided by the total number of samples is the LOSO-CV accuracy. 

The classification accuracy used for infants was the leave-one-experiment-block-out cross-validation (LOBO-CV) accuracy.  This alternative accuracy measure makes more effective use of the eye movement data when the sample size is very small.  In LOBO-CV, the samples belonging to one experiment block are sequestered, and the remaining samples are used to train the classifier.  The sequestered samples are then classified with the learned classifier, and the procedure is repeated for every block in the database.  The total number of correctly classified samples over the total number of samples is the LOBO-CV accuracy.  

%--------------------------------------------------------------------
\section{Results}
\subsection{Adult Experiment}\label{exp:adult1}
We first labeled the adult trials as category learner or non-learner.  This resulted in $728$ learning class samples and $1,256$ non-learning class samples for the learning phase, and $473$ learning class samples and $601$ non-learning class samples for the testing phase in the category A or B category learning condition.  There were $496$ learning class samples and $1,568$ non-learning class samples for the learning phase, and $323$ learning class samples and $717$ non-learning class samples for the testing phase in the category C or D category learning condition.  The indeterminate samples were not used in any of the experiments.  We then extracted the eye tracking variables from each trial's gaze sequence.  Each labeled data sample resulted in a $182$-dimensional feature vector for the learning phase samples, and a $334$-dimensional feature vector for the testing phase samples.   

\begin{figure}
\centering
\includegraphics[width=0.15\textwidth]{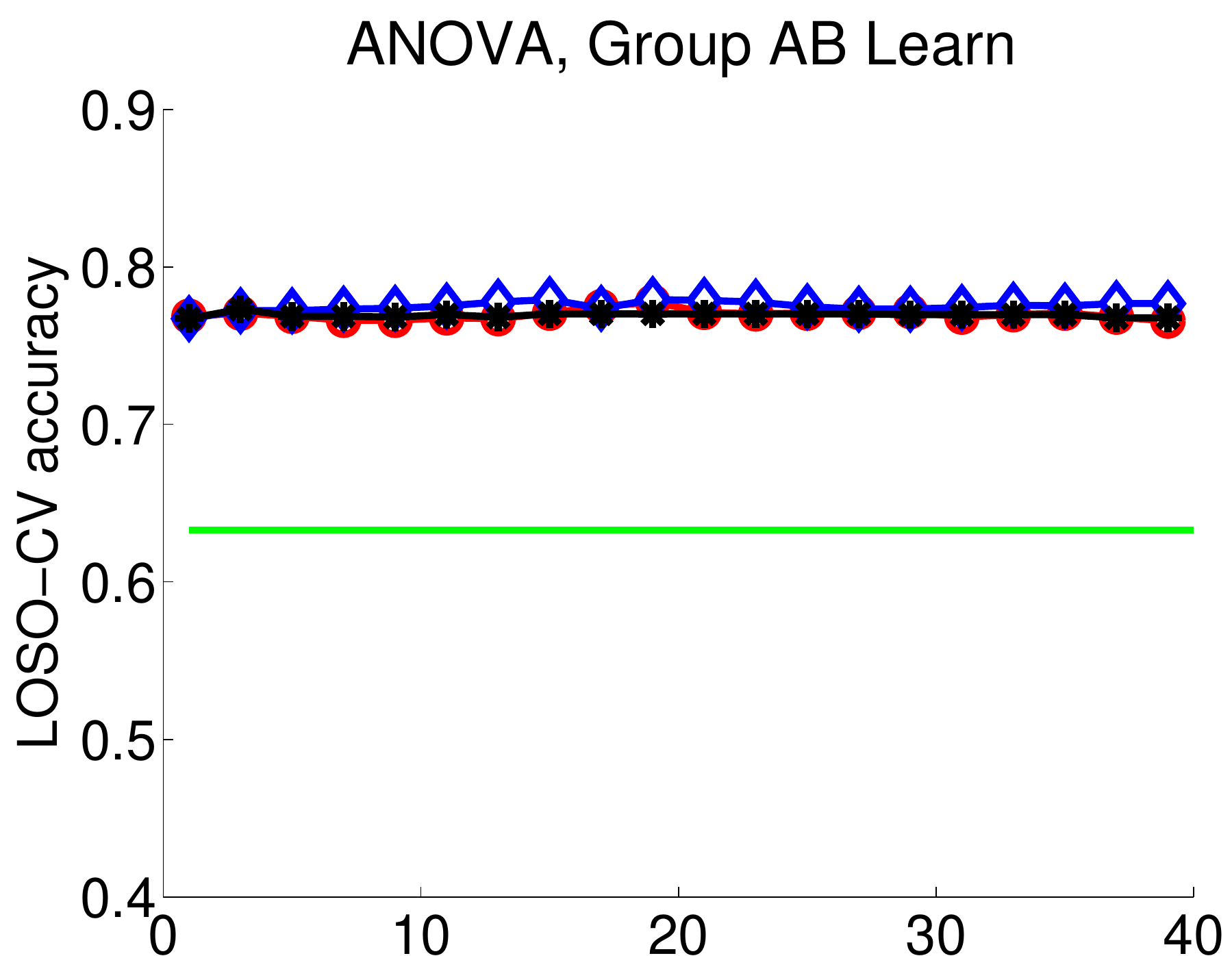}
\includegraphics[width=0.15\textwidth]{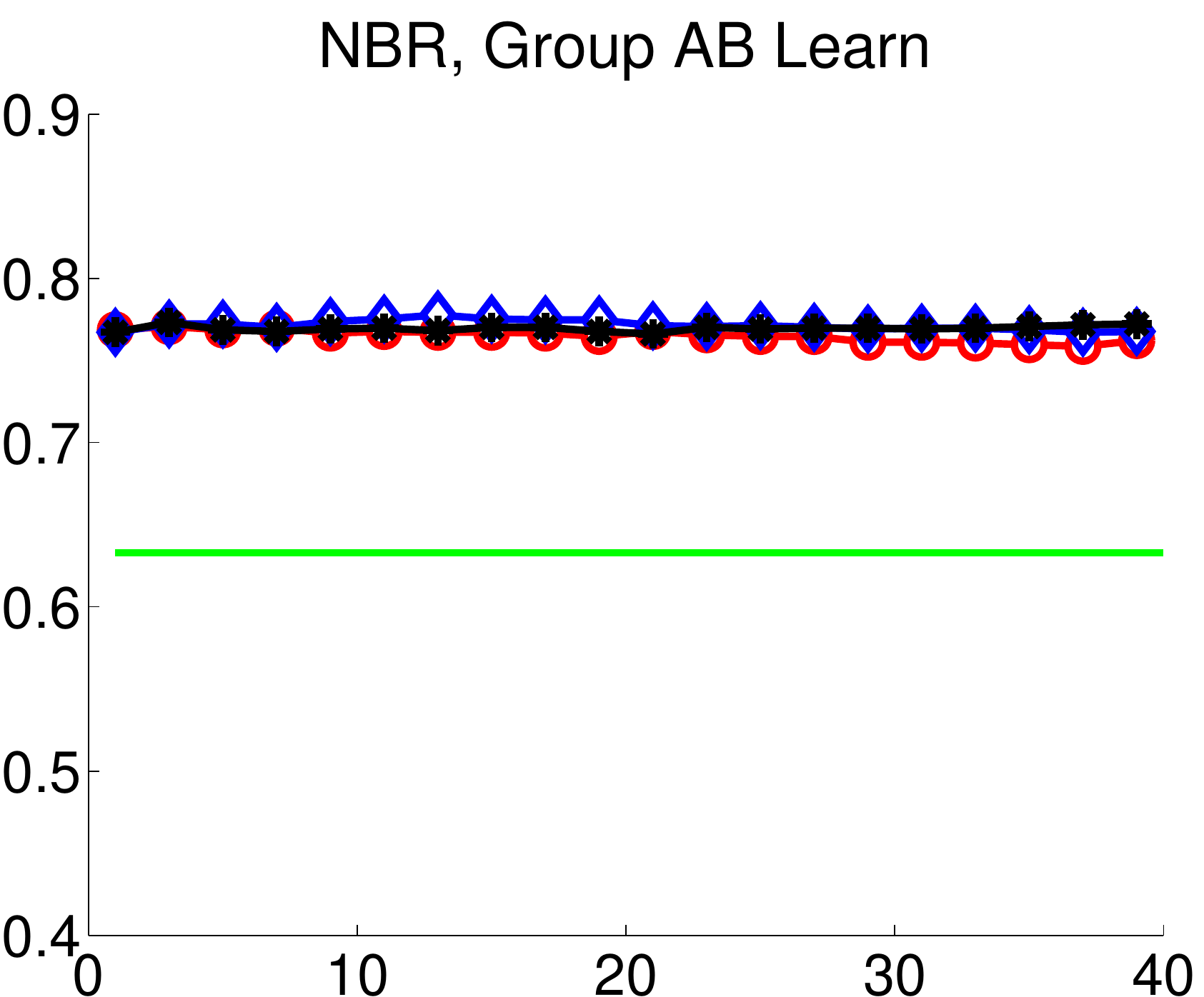}
\includegraphics[width=0.15\textwidth]{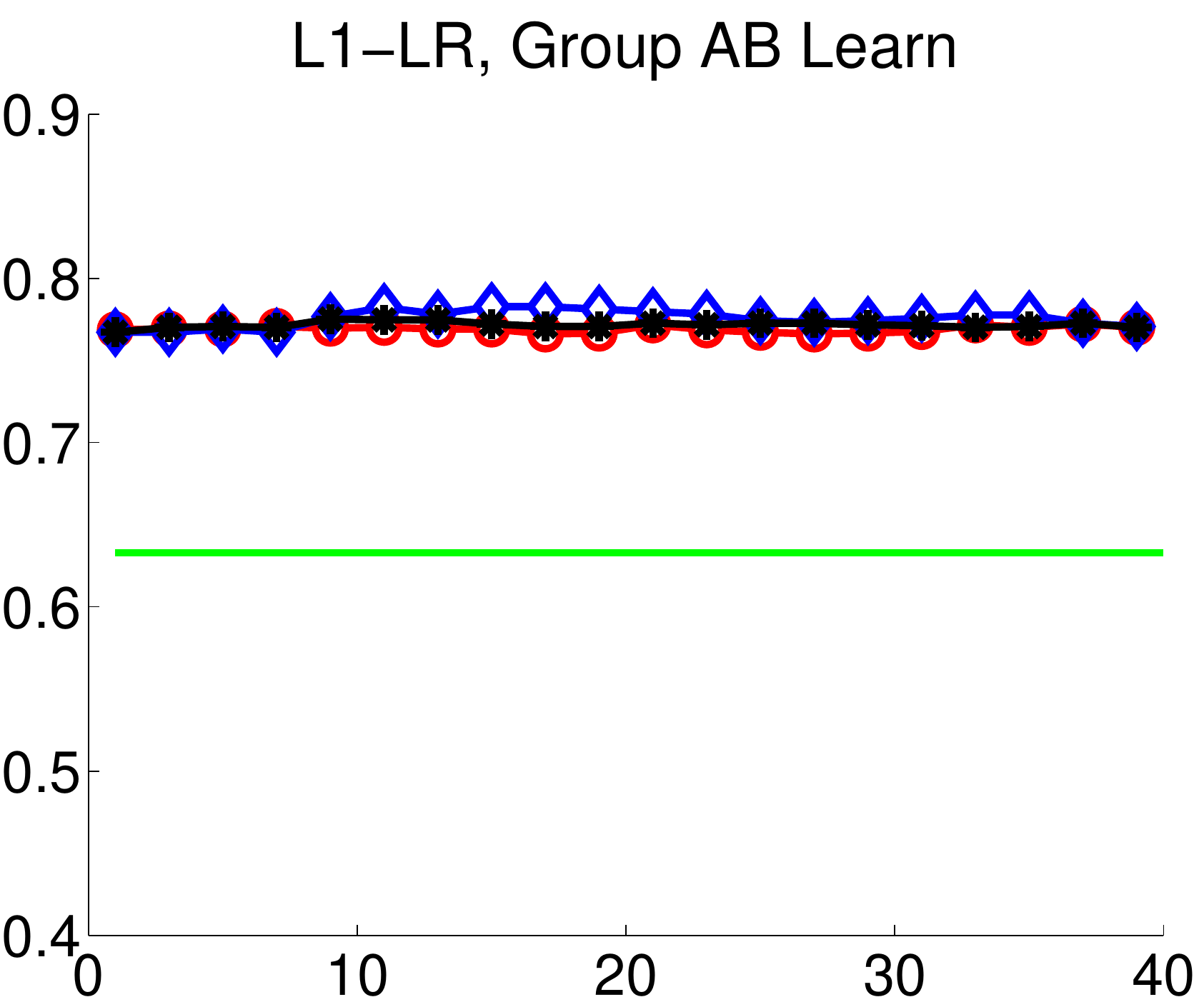}
\includegraphics[width=0.15\textwidth]{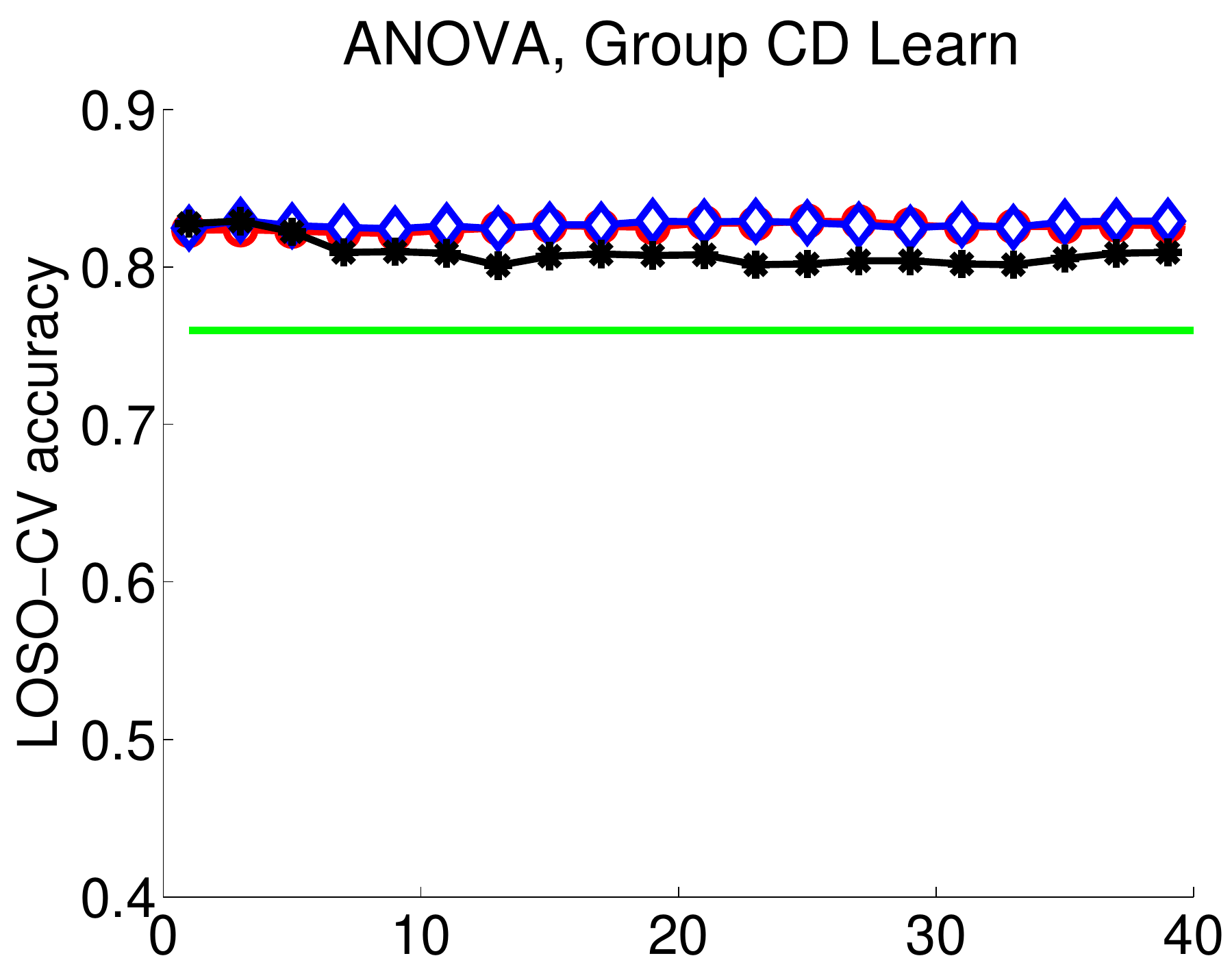}
\includegraphics[width=0.15\textwidth]{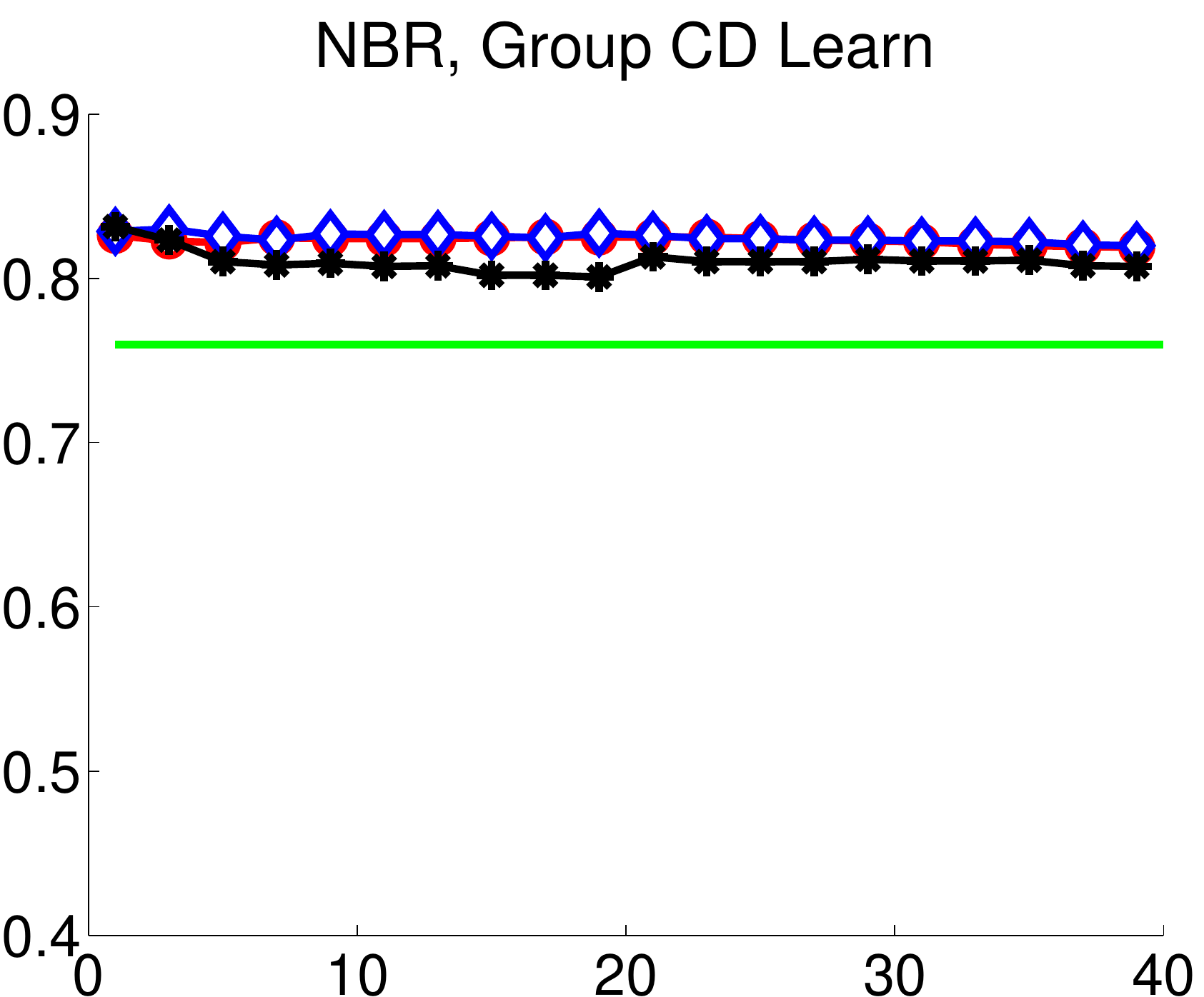}
\includegraphics[width=0.15\textwidth]{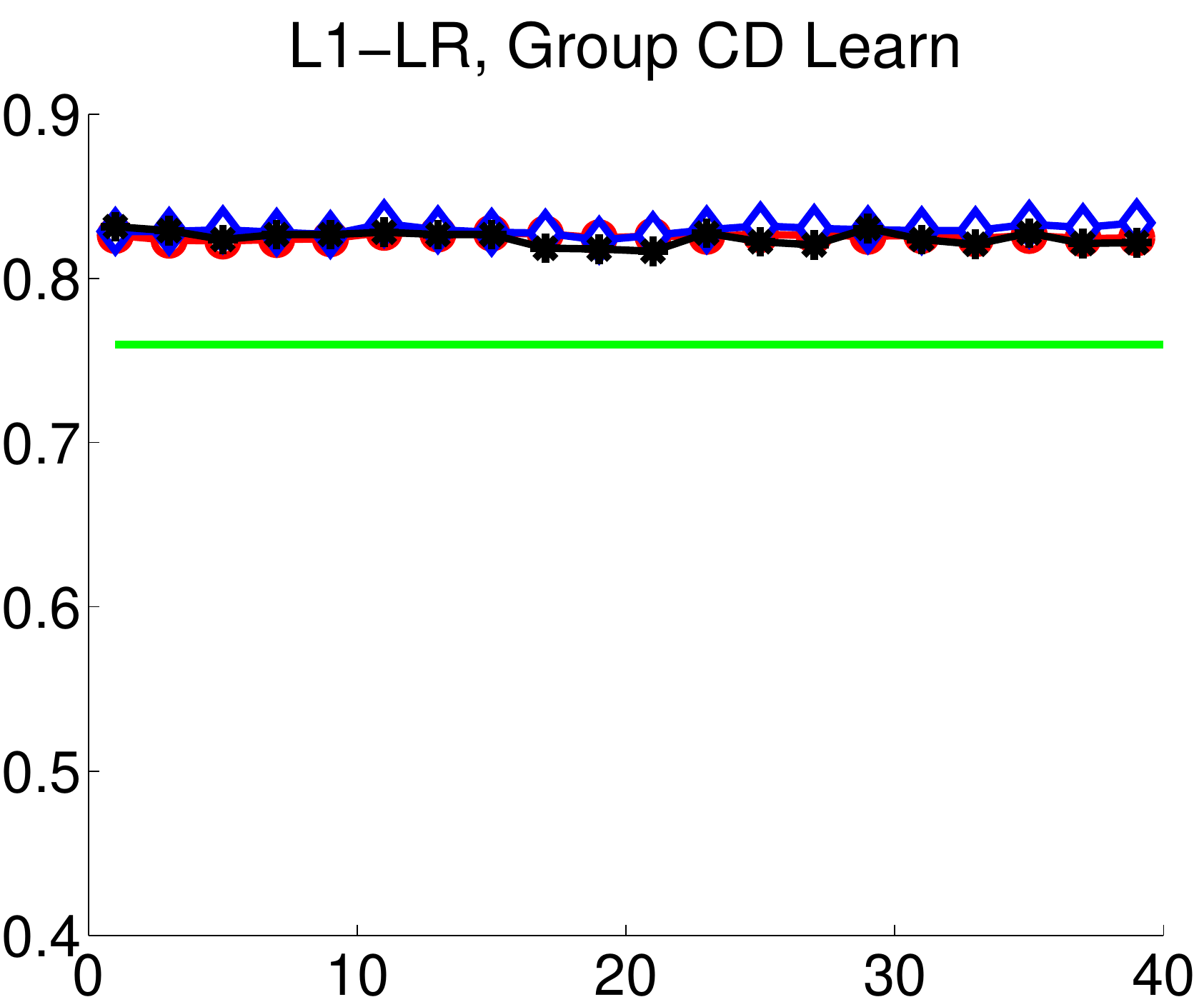}\\
\includegraphics[width=0.15\textwidth]{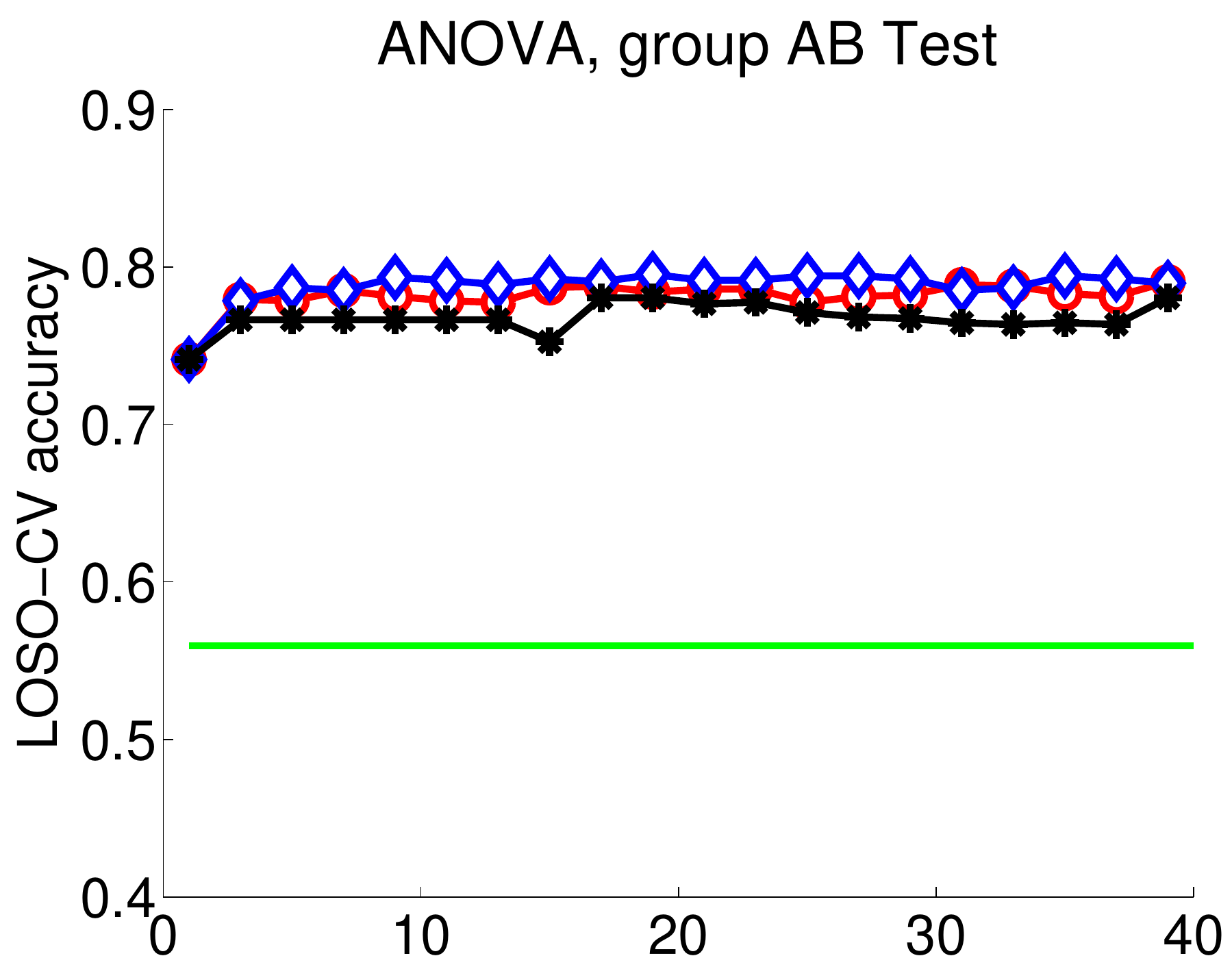}
\includegraphics[width=0.15\textwidth]{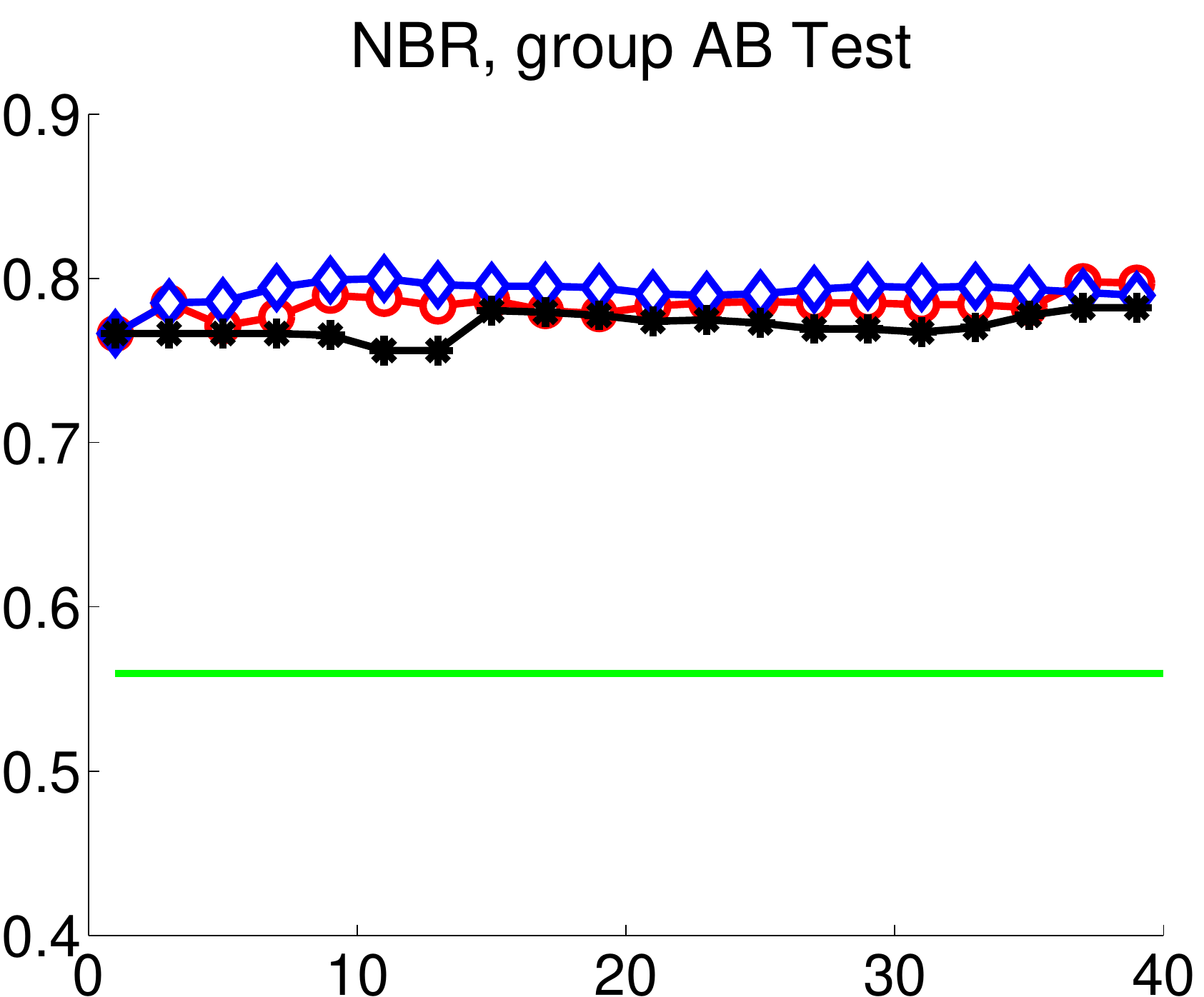}
\includegraphics[width=0.15\textwidth]{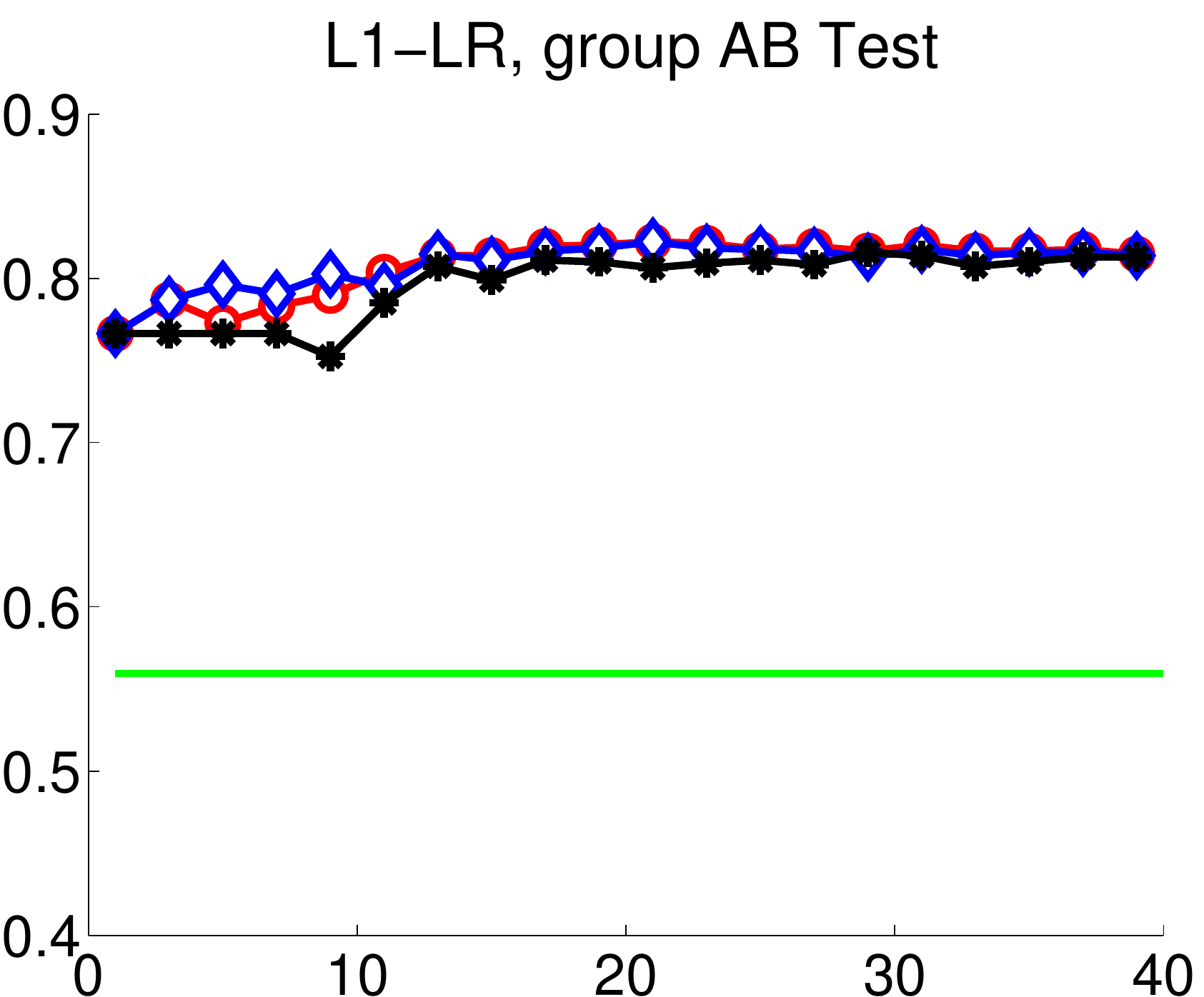}
\includegraphics[width=0.15\textwidth]{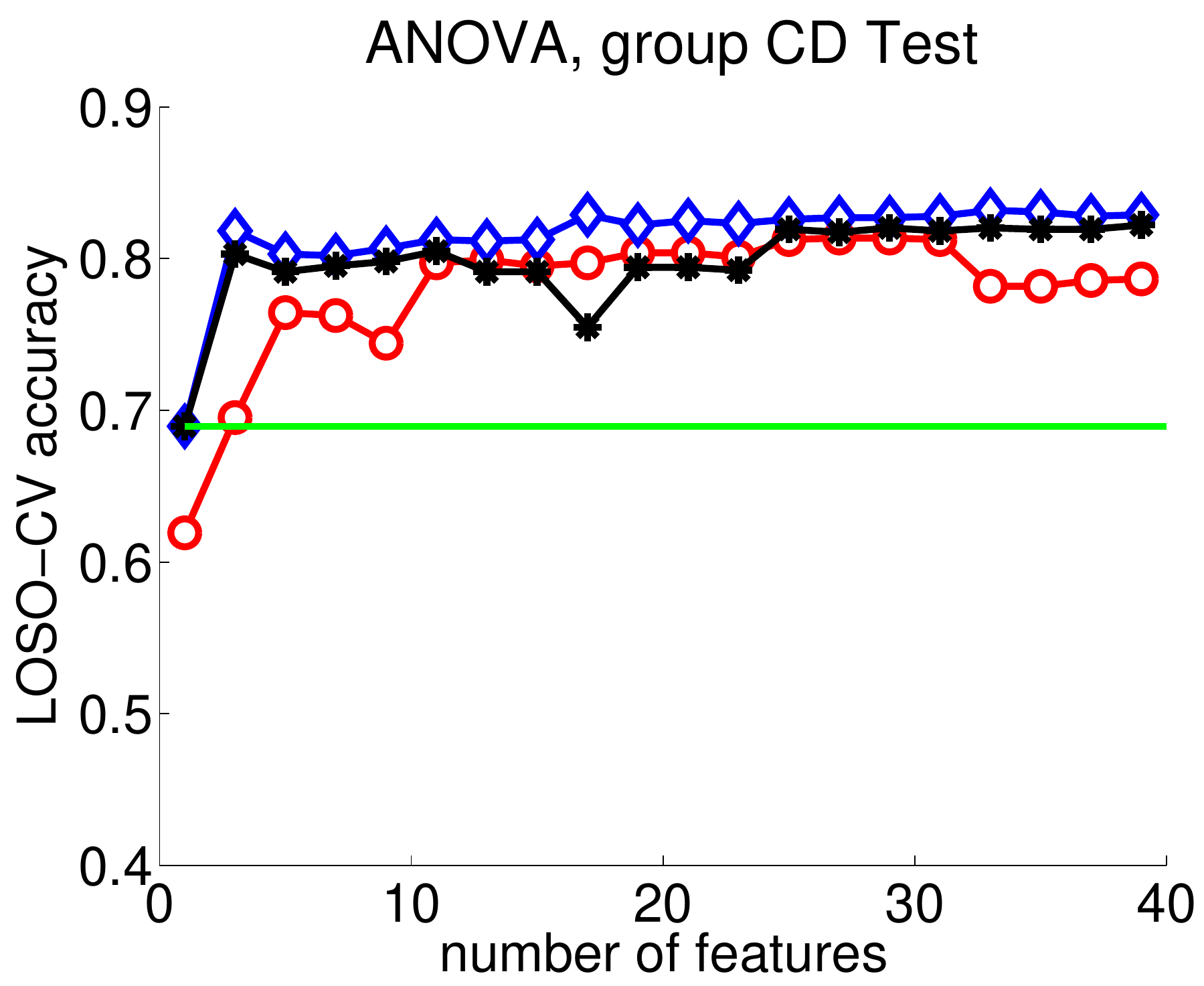}
\includegraphics[width=0.15\textwidth]{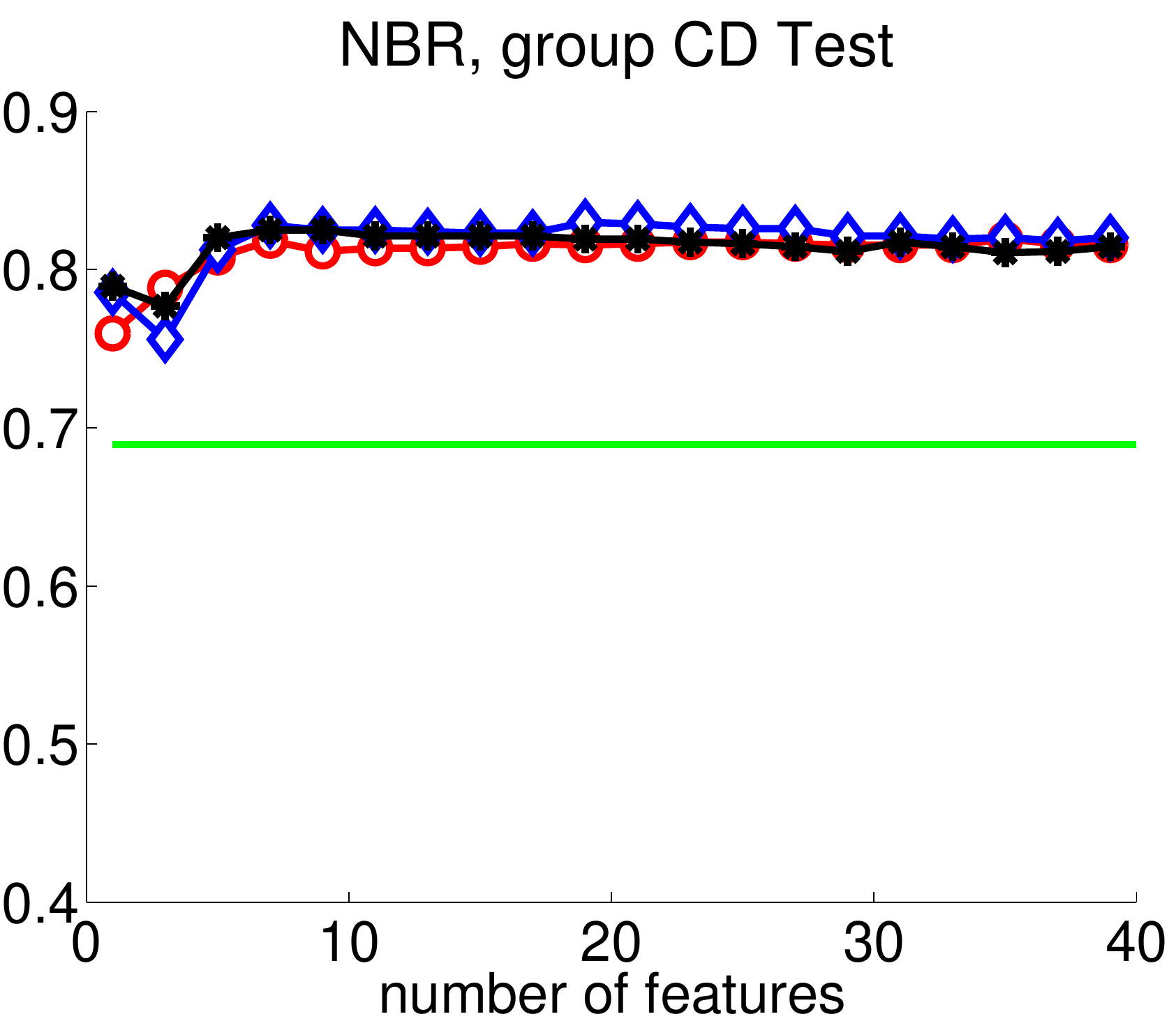}
\includegraphics[width=0.15\textwidth]{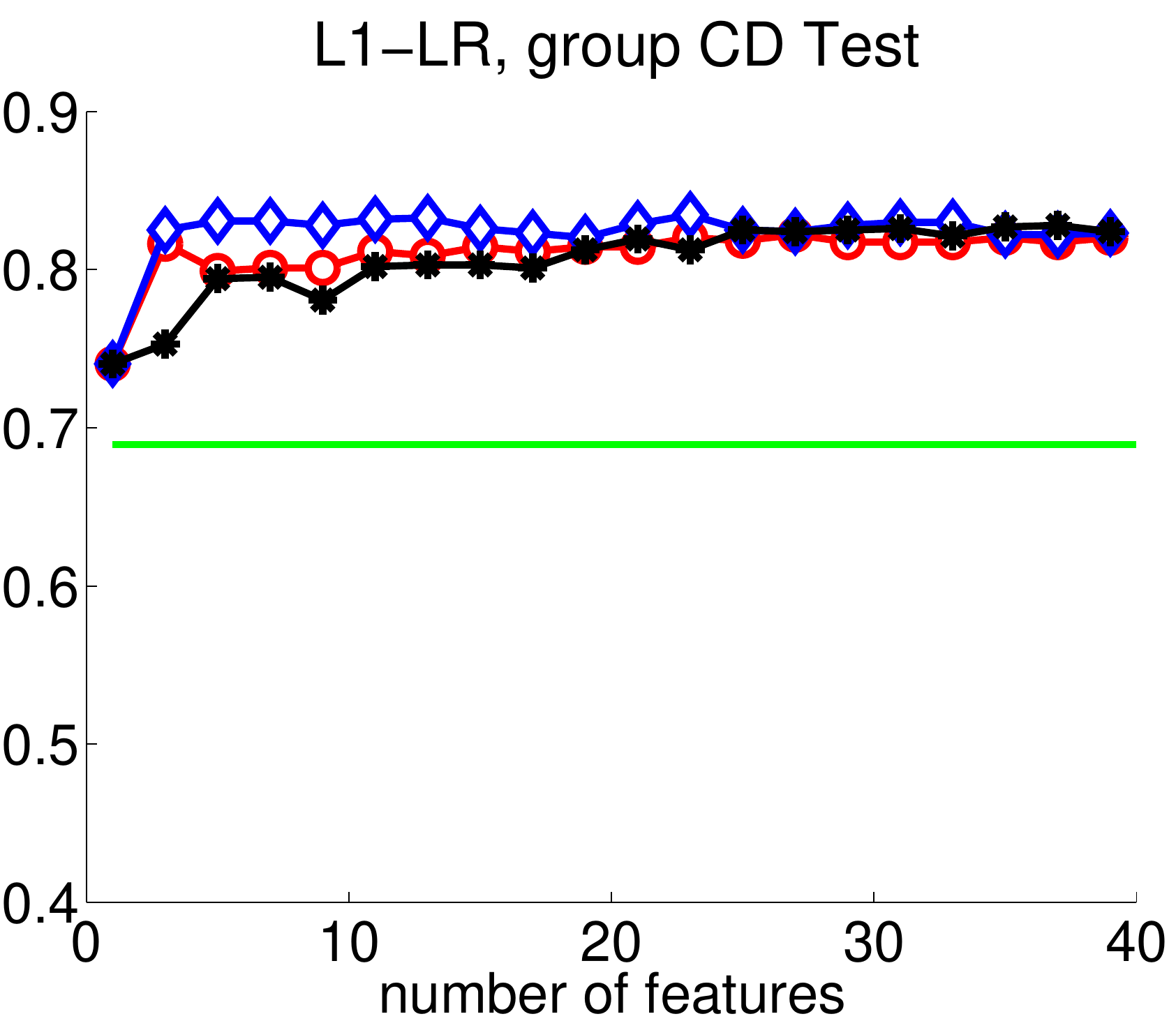}\\
\includegraphics[width=.4\textwidth]{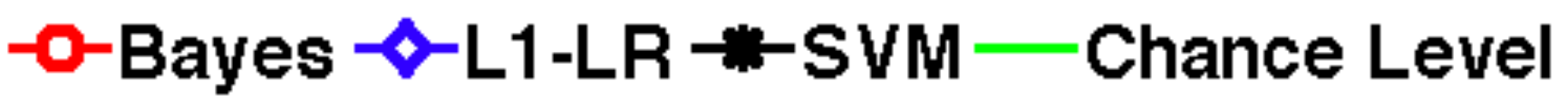}\\
\caption{Leave one subject out cross-validation accuracy for adult subjects as a function of the number of top ranked variables used for classification.  The first two rows show results for the learning phase of the experiments (categories AB and CD, respectively).  The last two rows show the results for the testing phase of the experiments (categories AB and CD, respectively).  ANOVA, NBR, and L1-LR correspond to ANOVA feature selection, Naive Bayes feature selection, and L1 penalized logistic regression feature selection, respectively.  AB and CD correspond to category object A or B and C or D respectively.  In almost all cases, the classification accuracy was near the maximum after including very few features and did not change much when including more.  Chance level is plotted as the accuracy resulting from classifying each sample as the most common class. }
\label{fig:adultEyeTrackResult}
\end{figure}

We applied the variable selection algorithms to identify the most important variables for separating learners from non-learners, and validated those variables using the three linear classifiers.  The LOSO-CV error is reported as a function of the number of top features used for classification in Fig. \ref{fig:adultEyeTrackResult}.  Recall that the features encode the eye tracking variables. The results show that a very small numbers of features yield a high classification rate, and including more features does not improve the accuracy.  

The stable performance beyond just a few features suggests that a small number of variables is sufficient for discriminating learners and non-learners.  The top five variables for ANOVA, NBR, and L1-LR are listed in Table \ref{table:adultTesting}.  We boldfaced the variables that were consistently ranked in the top five variables across both the category A or B and category C or D conditions and all feature selection algorithms.  We underlined variables that were consistently ranked in the top five variables by at least two of the three features selection algorithms and across category A or B and category C or D conditions.  Note that AOI $4$ for the category A or B condition is equivalent to AOI $6$ in the category C or D condition.  The consistent top variables in the learning condition were \emph{latency to a fixation at the relevant AOI}, \emph{density of fixations at the relevant AOI}, and \emph{first fixation at the relevant AOI}. The top variables in the testing condition were \emph{first, second, and third fixations}.

%-----------------------------------------
\begin{table}
%\centering
\scriptsize
%\tiny
% learning condition
\begin{tabular}{ r l }
\begin{sideways}\hspace{-.2cm}A or B\end{sideways} &\hspace{-.3cm} 
\begin{tabular}{  l  l  l }
\emph{ Learning condition} & & \\ \hline
\hline
\hline
~~~~ANOVA & NBR & L1-LR \\  \hline
1. {\bf Lat to rel AOI fix } & {\bf  Lat to rel AOI fix } & {\bf  Den of fix at AOI 4}\\
2. {\bf Den of fix at AOI 4} & {\bf  Den of fix at AOI 4} &  {\bf  Lat to rel AOI fix }  \\ 
3. AOI 4, DHB 2  &  AOI 4, DHB 2   &   $2^{nd}$ fix at AOI 4 \\
4. $2^{nd}$ fix at AOI 4 &  $2^{nd}$ fix at AOI 4  & $5^{th}$ fix at AOI 4\\ 
5. \underline{$1^{st}$ fix at AOI 4 } & \underline{ $1^{st}$ fix at AOI 4}  & $3^{rd}$ fix at rel AOI \\
\hline 
\end{tabular}
\end{tabular} %end AB tabular

\begin{tabular}{ r  l  }
\begin{sideways}\hspace{0cm}C or D\end{sideways} & \hspace{-.3cm} 
\begin{tabular}{ l  l  l }  
\hline
1. {\bf  Lat to rel AOI fix} & {\bf  Den of fix at AOI 6}  &{\bf  Den of fix at AOI 6}   \\
2. {\bf  Den of fix at AOI 6} & {\bf Lat to rel AOI fix} & {\bf  Lat to rel AOI fix} \\ 
3. AOI 6, DHB 5  &  AOI 6, DHB 2   & \underline{$1^{st}$ fix at rel AOI}\\
4. \underline{$1^{st}$ fix at AOI 6}  & \underline{$1^{st}$ fix at AOI 6 }&   $1^{st}$ sac to rel AOI   \\ 
5. $1^{st}$ fix at non-rel AOI & $2^{nd}$ fix at AOI 6  & Den of fix at AOI 1 \\
\hline  
\hline
\hline
\end{tabular}
\end{tabular} %end CD tabular

%testing condition
\begin{tabular}{ r l }
\begin{sideways}\hspace{-.2cm}A or B\end{sideways} &\hspace{-.3cm} 
\begin{tabular}{ l  l  l }
\emph{Testing condition} & & \\ \hline
\hline
\hline
~~~~ANOVA & NBR & L1-LR  \\  \hline
1.  \underline{$3^{rd}$ fix at non-rel AOI }& \underline{ $2^{nd}$ fix at non-rel AOI}  & \underline{ $2^{nd}$ fix at non-rel AOI } \\
2. \underline{ $2^{nd}$ fix at non-rel AOI } & \underline{ $2^{nd}$ sac to non-rel AOI }  & \underline{$3^{rd}$ fix at non-rel AOI }\\ 
3. \underline{ $2^{nd}$ sac to non-rel AOI} &\underline{ $1^{st}$ fix at non-rel AOI } & \underline{  $1^{st}$ fix at non-rel AOI } \\
4. \underline{ $1^{st}$ fix at non-rel AOI} &  Duration of $3^{rd}$ fix   &  \underline{ $2^{nd}$ sac to non-rel AOI} \\ 
5. Number AOIs fixated &  \underline{ $3^{rd}$ fix at non-rel AOI }  &  $1^{st}$ sac to non-rel AOI\\ 
\hline
\end{tabular}
\end{tabular}%end A or B tabular

\begin{tabular}{ r l }
\begin{sideways}\hspace{0cm}C or D\end{sideways} &\hspace{-.3cm} 
\begin{tabular}{ l  l  l }
\hline
1. $4^{th}$ fix at non-rel AOI &  Rel AOI fix density& \underline{ $2^{nd}$ sac to non-rel AOI } \\
2. \underline{$3^{rd}$ fix at non-rel AOI }&\underline{ $1^{st}$ fix at non-rel AOI}  & \underline{ $1^{st}$ fix at non-rel AOI } \\ 
3. \underline{$2^{nd}$ sac to non-rel AOI } & Den of fix at AOI 13 & Rel AOI fix density \\
4. Number AOIs fixated & $1^{st}$ fix at rel AOI & \underline{ $2^{nd}$ fix at non-rel AOI}\\ 
5. $3^{rd}$ sac to non-rel AOI  & \underline{ $2^{nd}$ fix at non-rel AOI } & \underline{ $3^{rd}$ fix at non-rel AOI} \\
\hline
\hline
\hline
\end{tabular}
\end{tabular}%end C  or C tabular
\caption{Adult Experiment: The following variables were determined most relevant during the category learning and category discrimination phases of the adult experiment.  The bold face entries show variables that were consistently determined most relevant using all feature selection algorithms and on two separate category object conditions.  The underlined entries show variables that were determined most relevant by at least two feature selection algorithms and across both category conditions.
ANOVA, NBR, and L1-LR correspond to the different feature selection algorithms. AOI 4 is relevant in the category A or B condition, and corresponds to AOI 6 in the category C or D condition. We use the following shorthand convention: fixation (fix), saccade (sac), relevant (rel), density (den), latency (lat), distance histogram bin (DHB).   }
\label{table:adultTesting}
\end{table}
%-----------------------------------------

\subsection{Infant Experiment}\label{exp:infant1} 
We first labeled the infant trials as category learner or non-learner.  This amounted to $135$ learning class samples and $137$ non-learning class samples for the learning phase, and $40$ learning class samples and $40$ non-learning class samples for the testing phase in the category A or B category learning condition. 
The C or D category learning condition resulted in $139$ learning class samples and $127$ non-learning class samples for the learning phase, and $40$ learning class samples and $40$ non-learning class samples for the testing phase.  As in the adult experiment, the indeterminate samples were not used.  After labeling the data and extracting the variables from each gaze sequence, each sample resulted in a $334$-dimensional feature vector for both the learning and testing phase samples.

\begin{figure}
\centering
\includegraphics[width=0.15\textwidth]{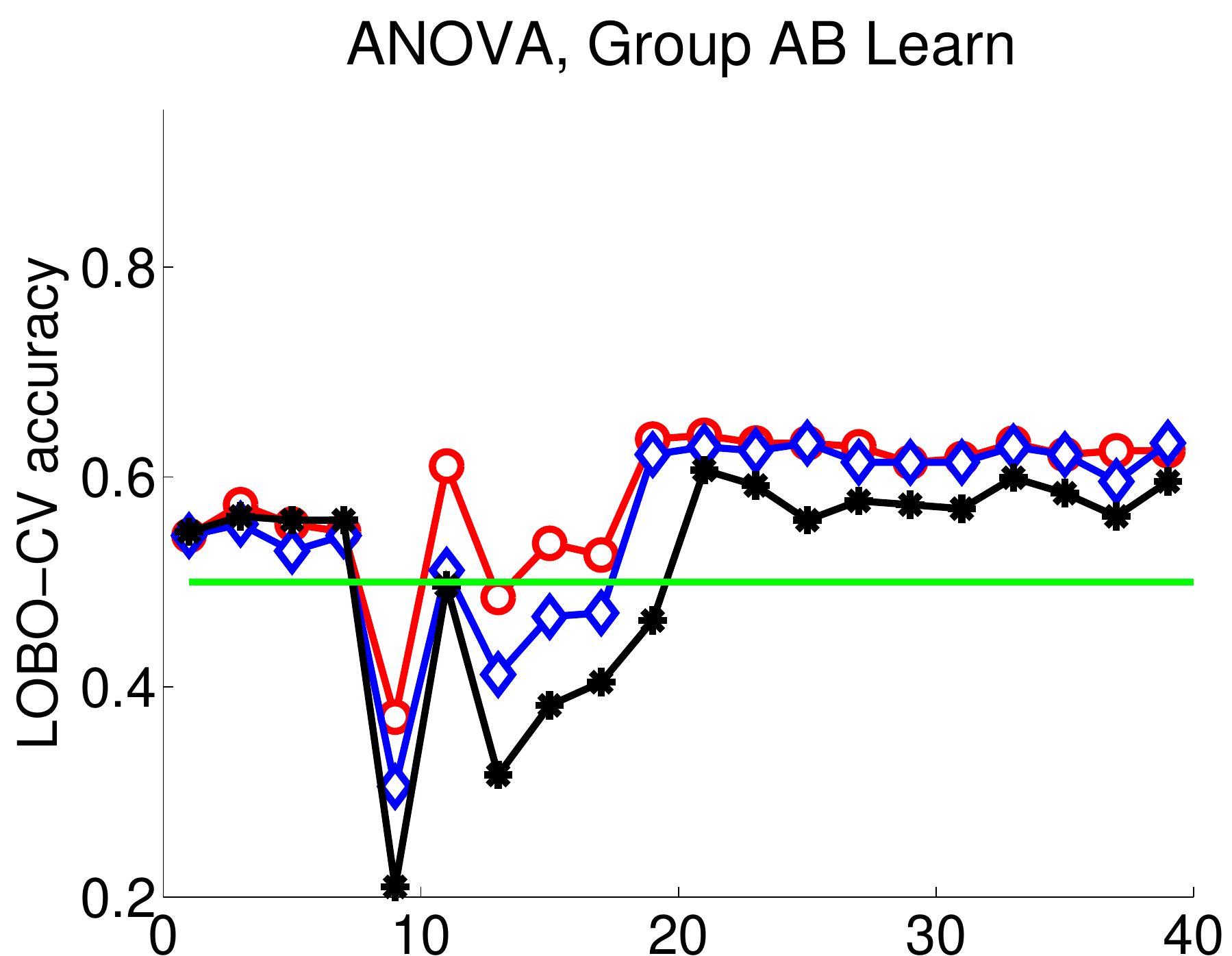}
\includegraphics[width=0.15\textwidth]{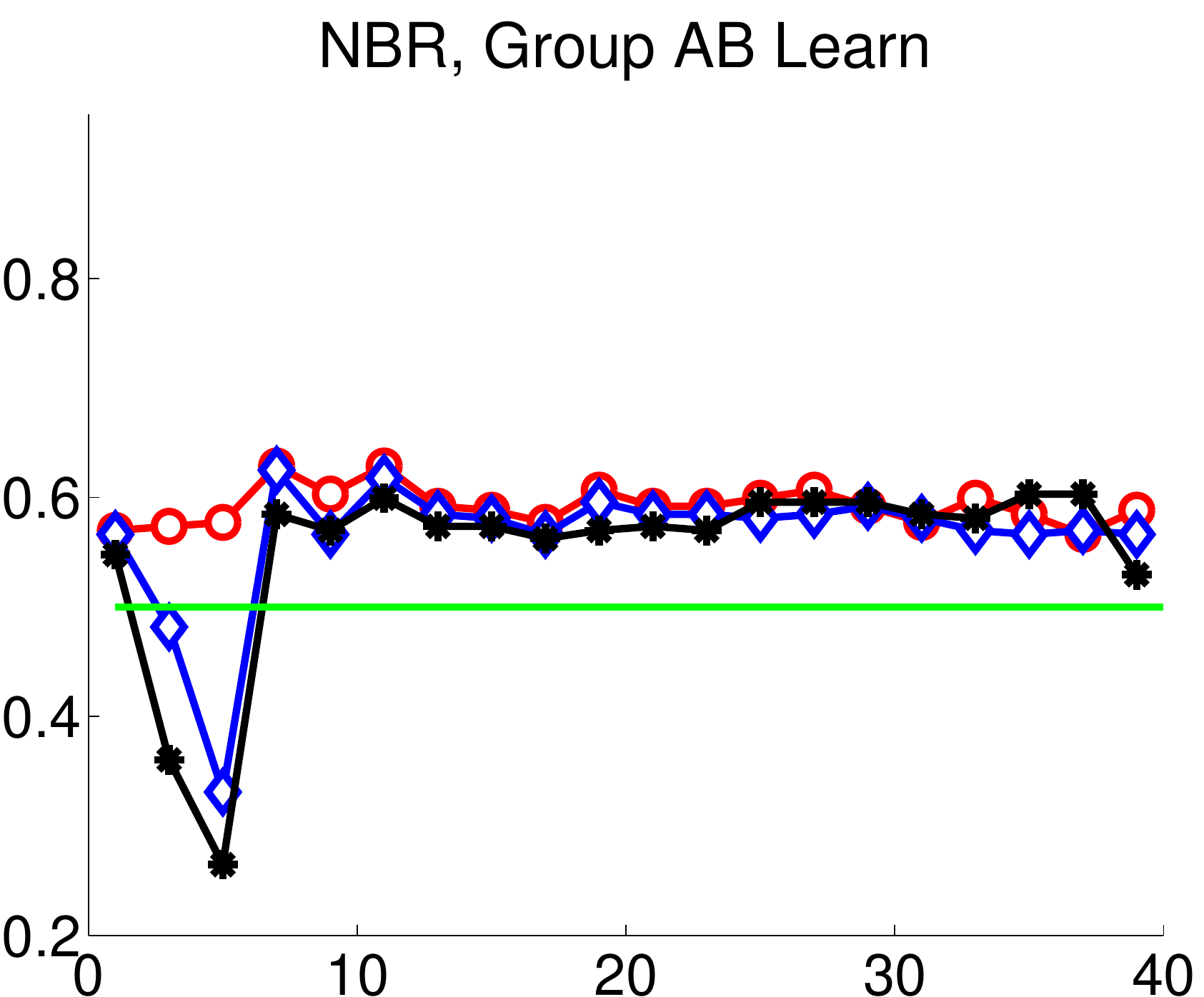}
\includegraphics[width=0.15\textwidth]{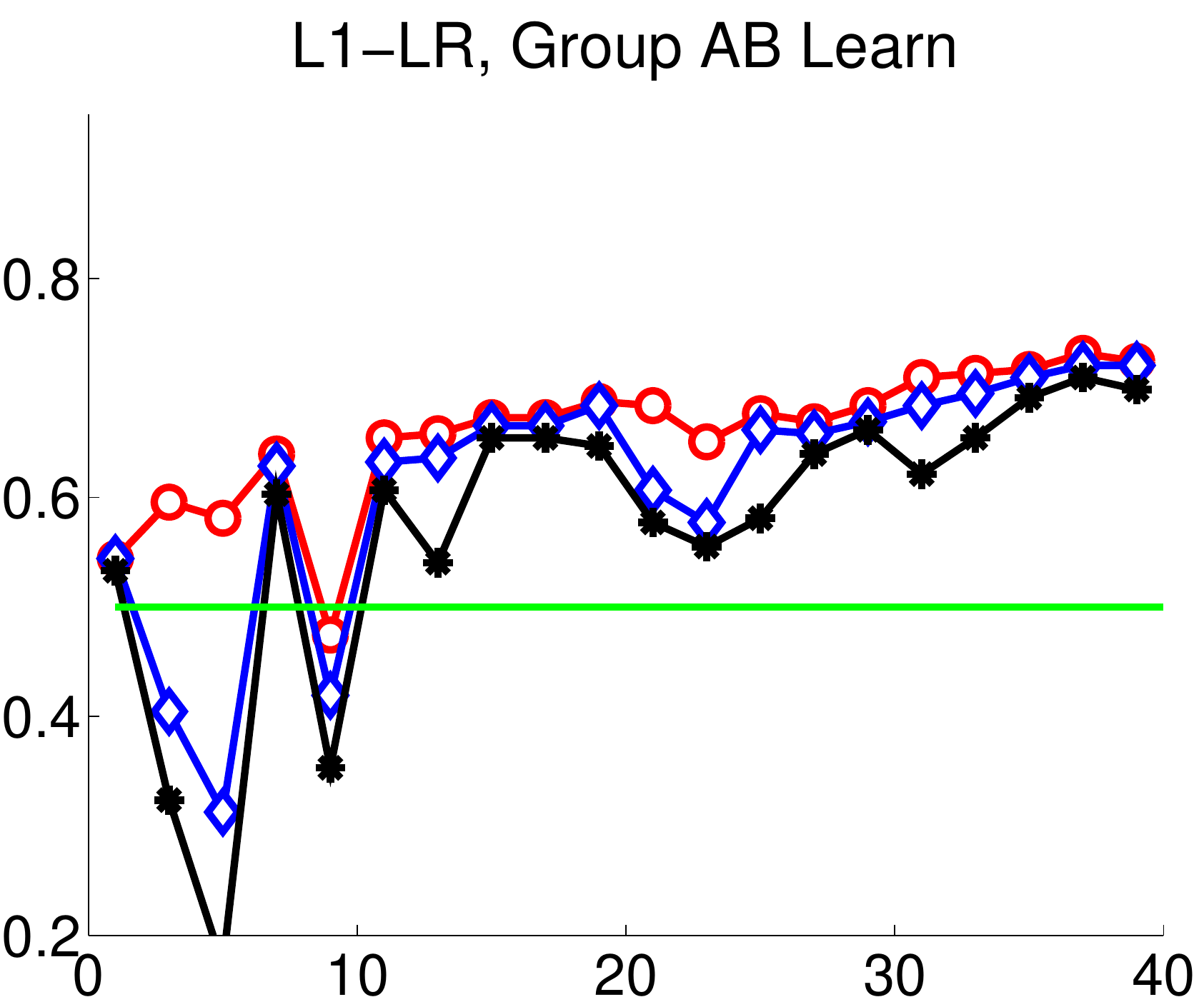}
\includegraphics[width=0.15\textwidth]{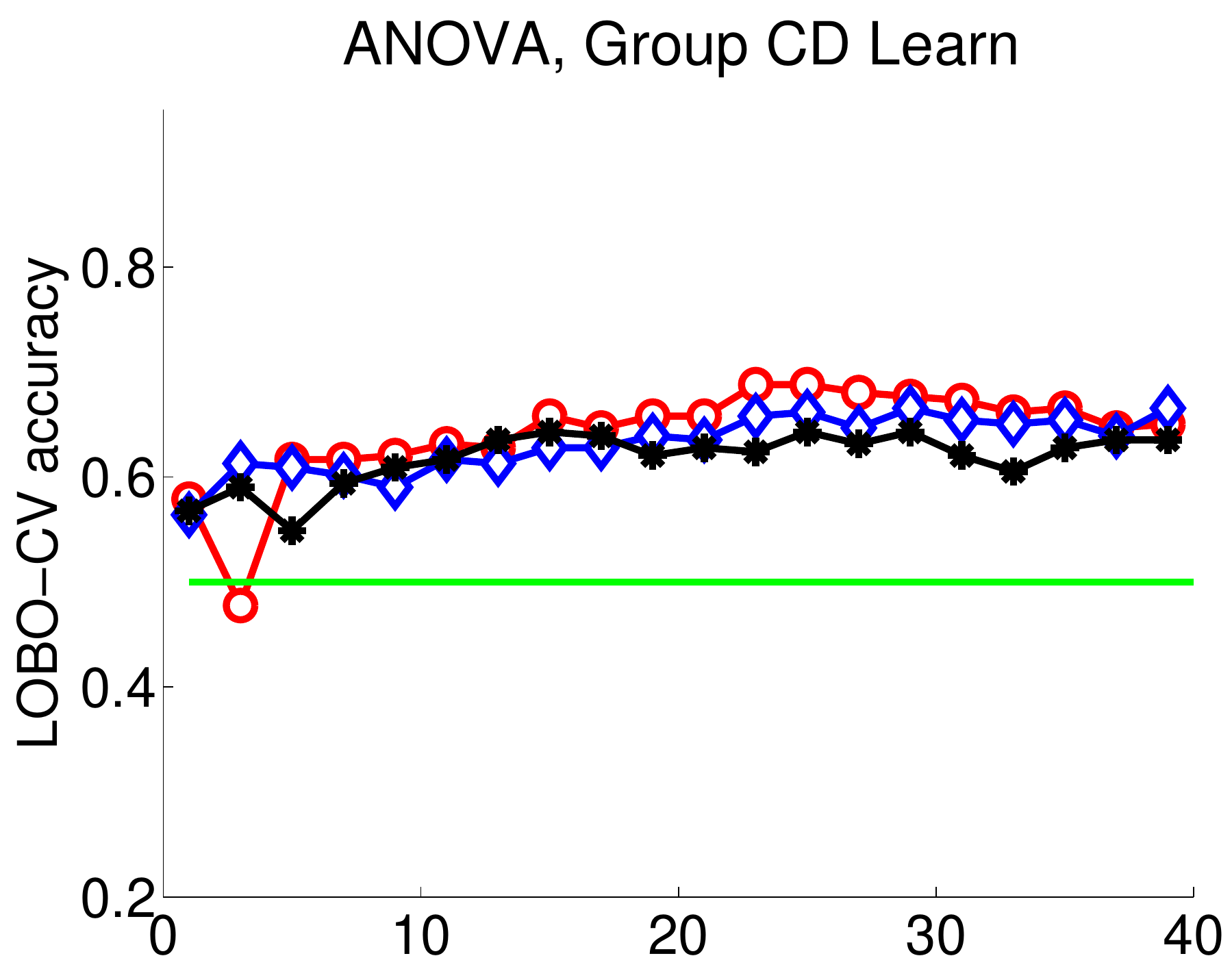}
\includegraphics[width=0.15\textwidth]{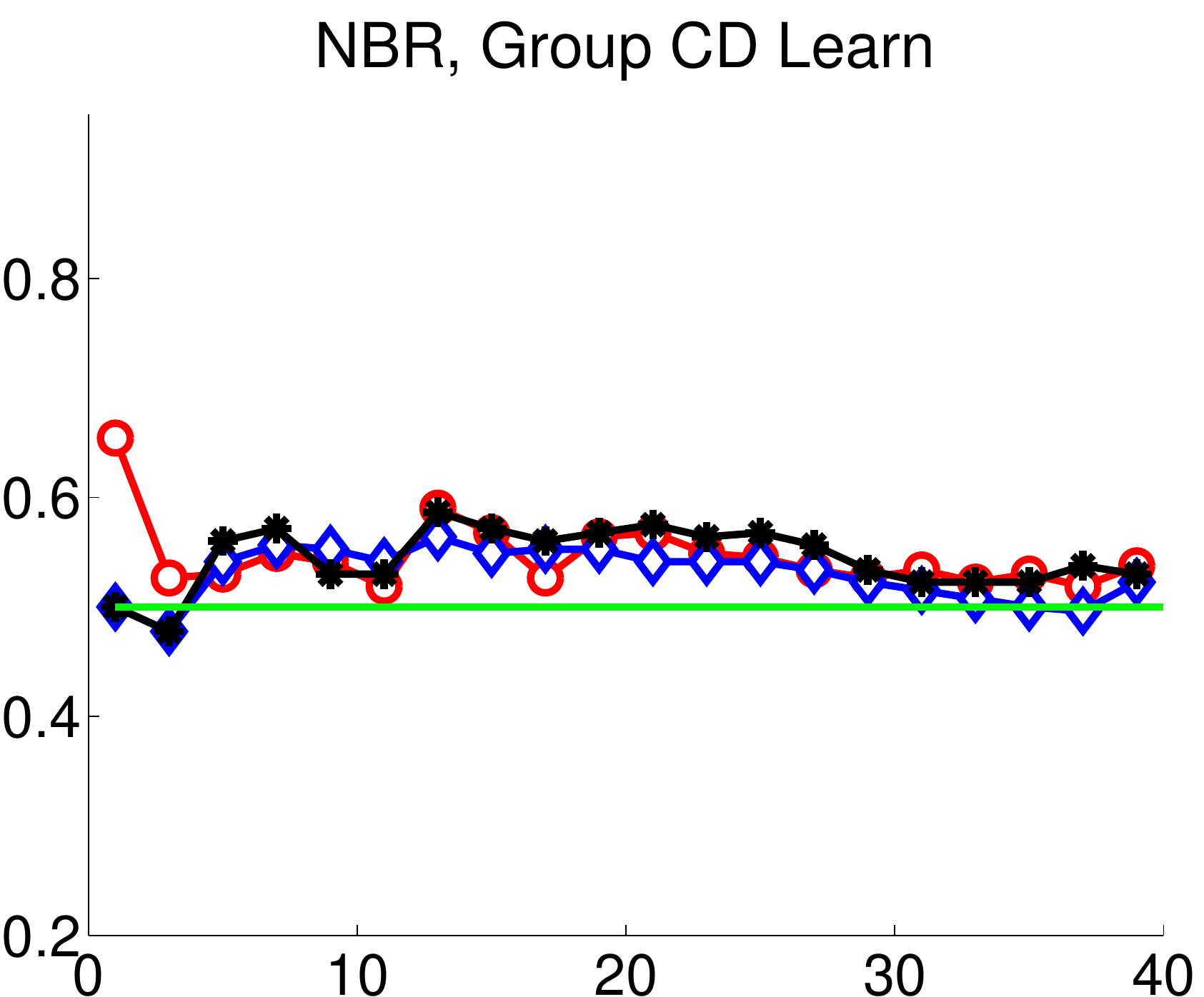}
\includegraphics[width=0.15\textwidth]{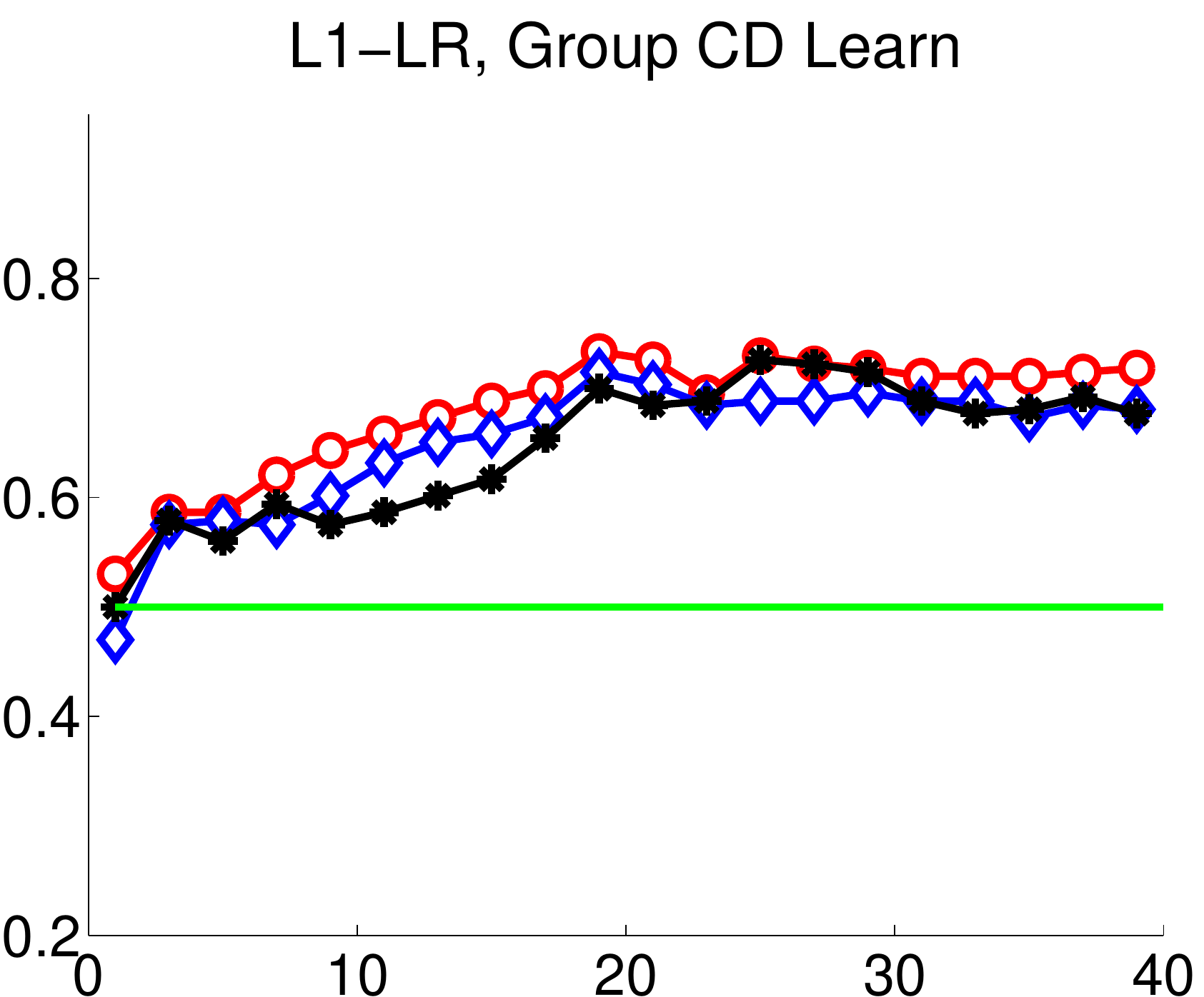}\\
\includegraphics[width=0.15\textwidth]{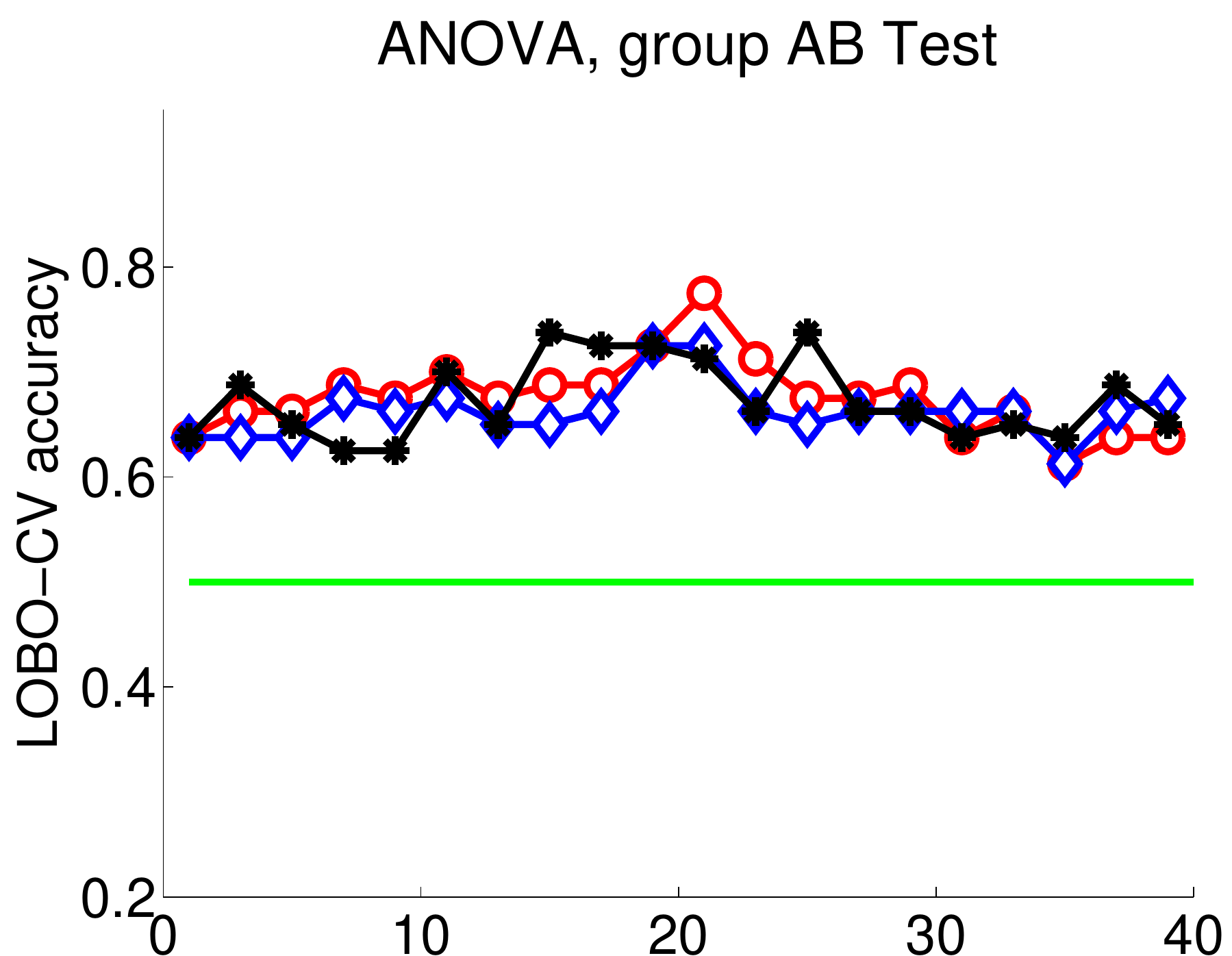}
\includegraphics[width=0.15\textwidth]{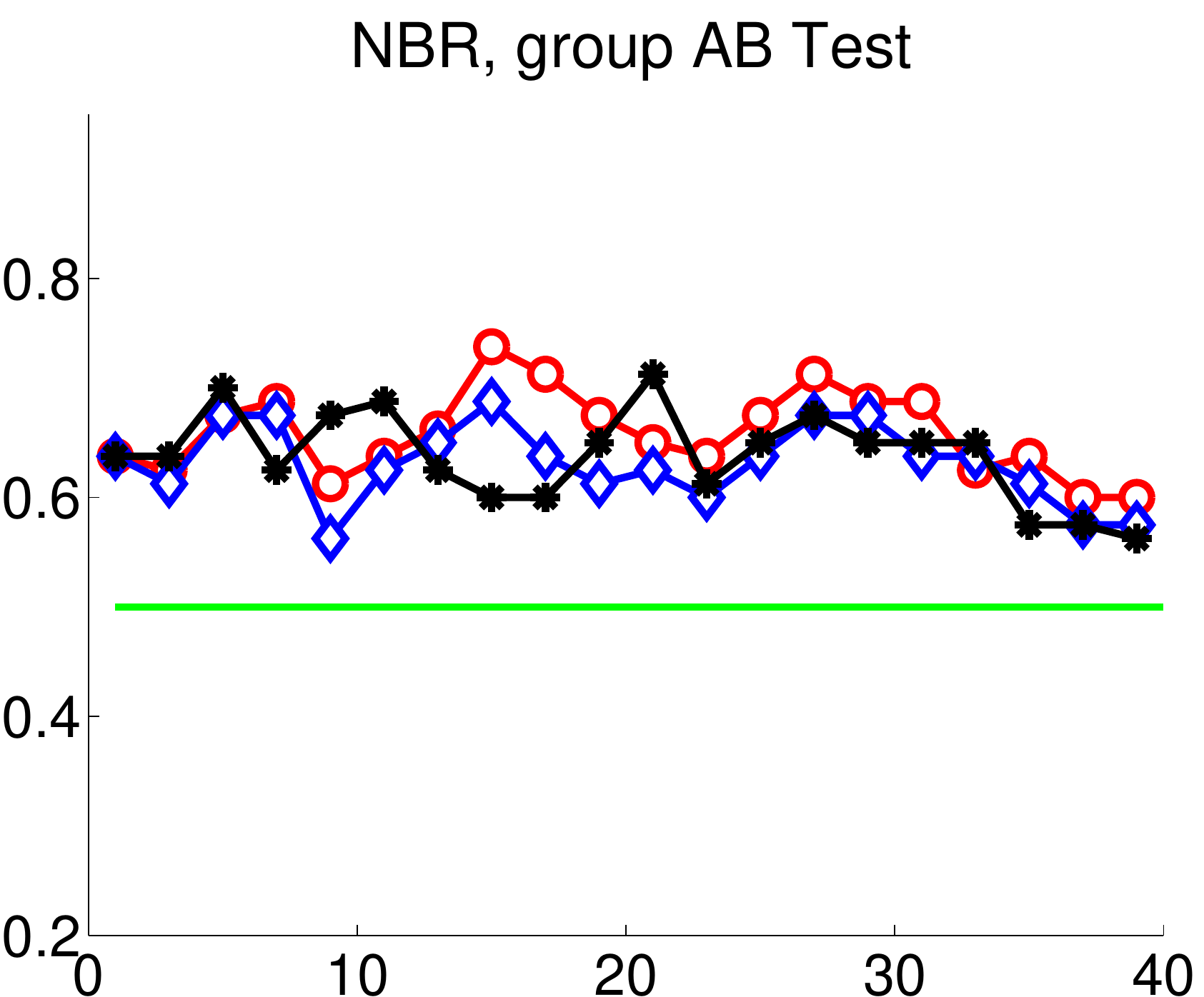}
\includegraphics[width=0.15\textwidth]{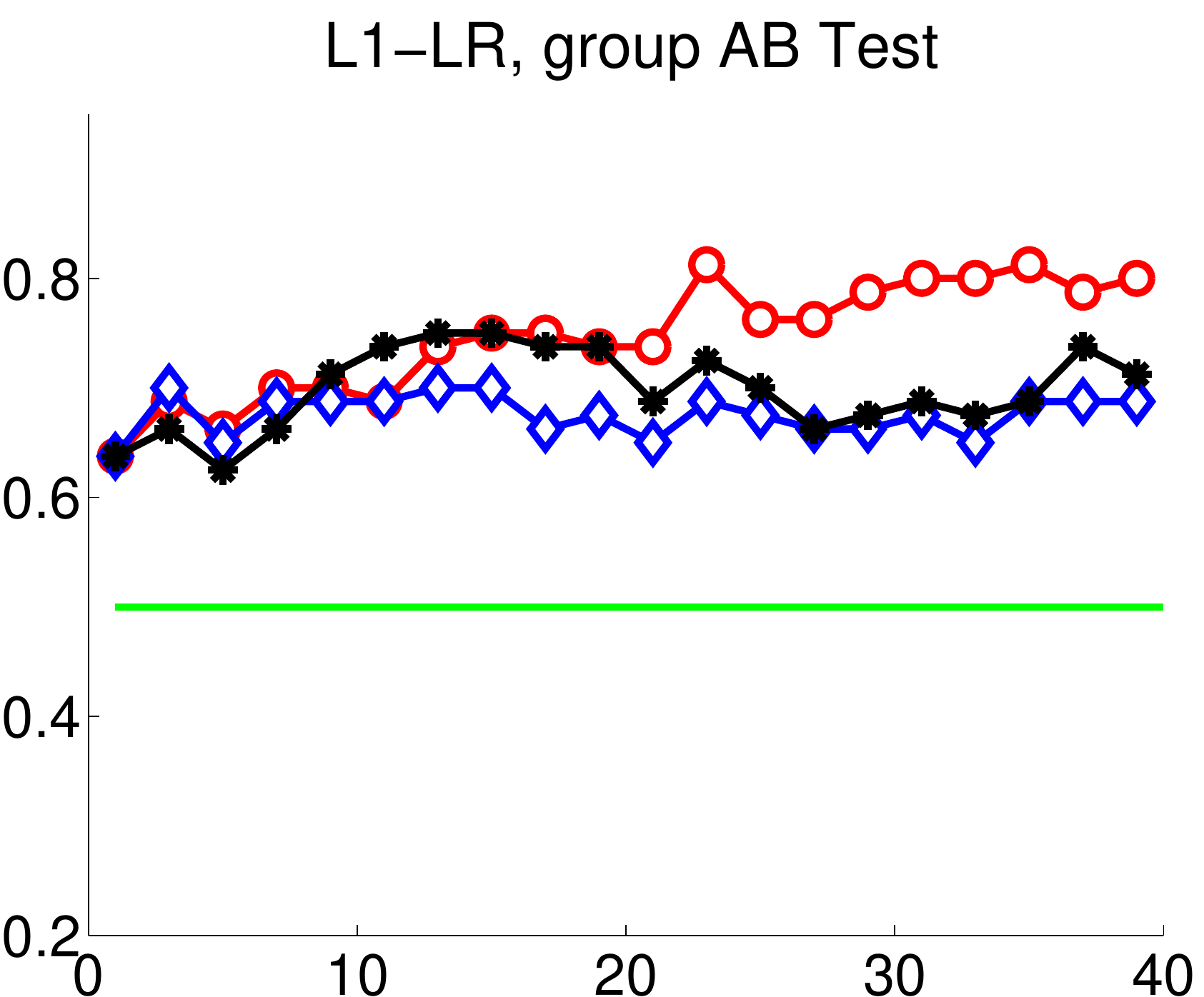}
\includegraphics[width=0.15\textwidth]{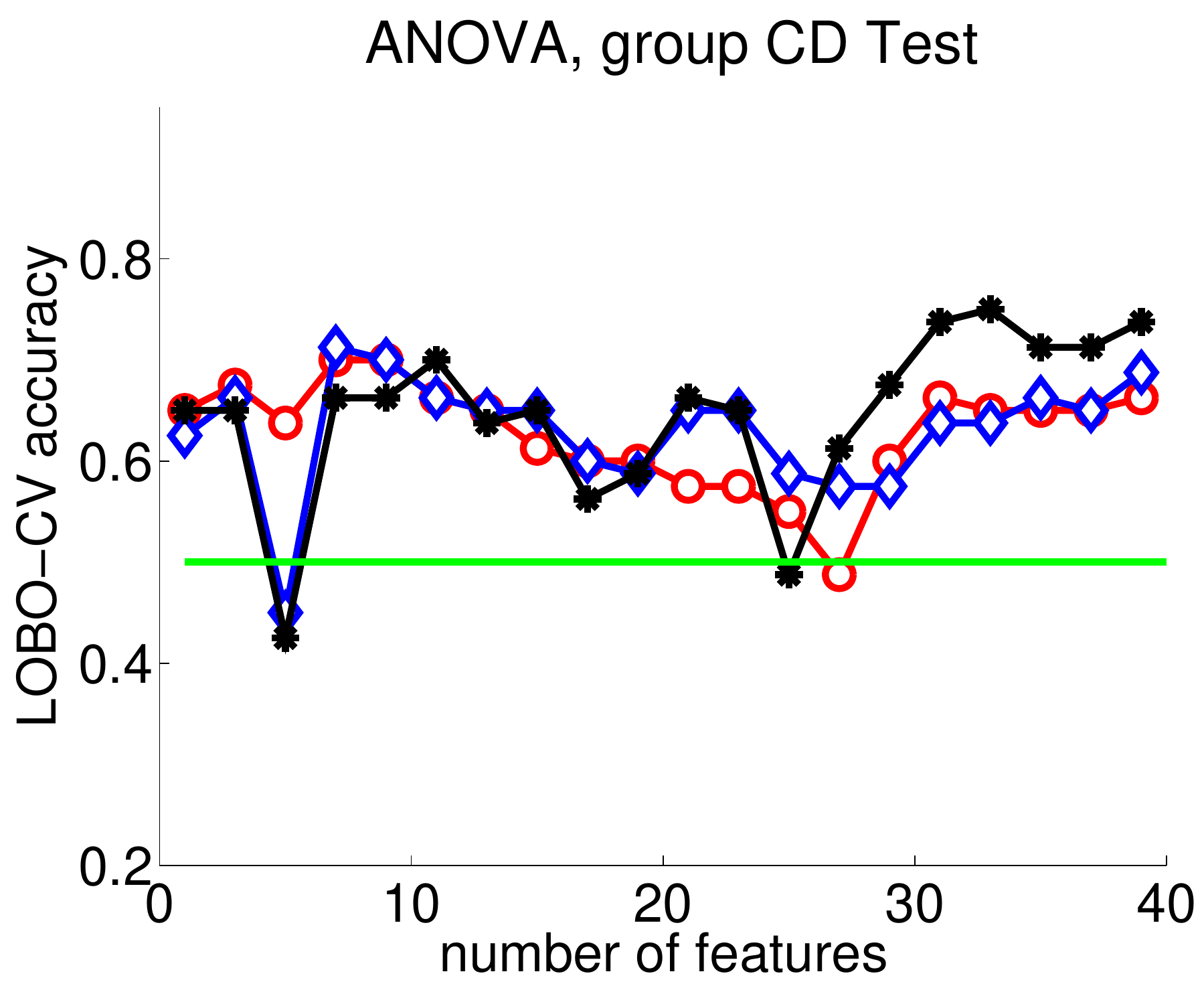}
\includegraphics[width=0.15\textwidth]{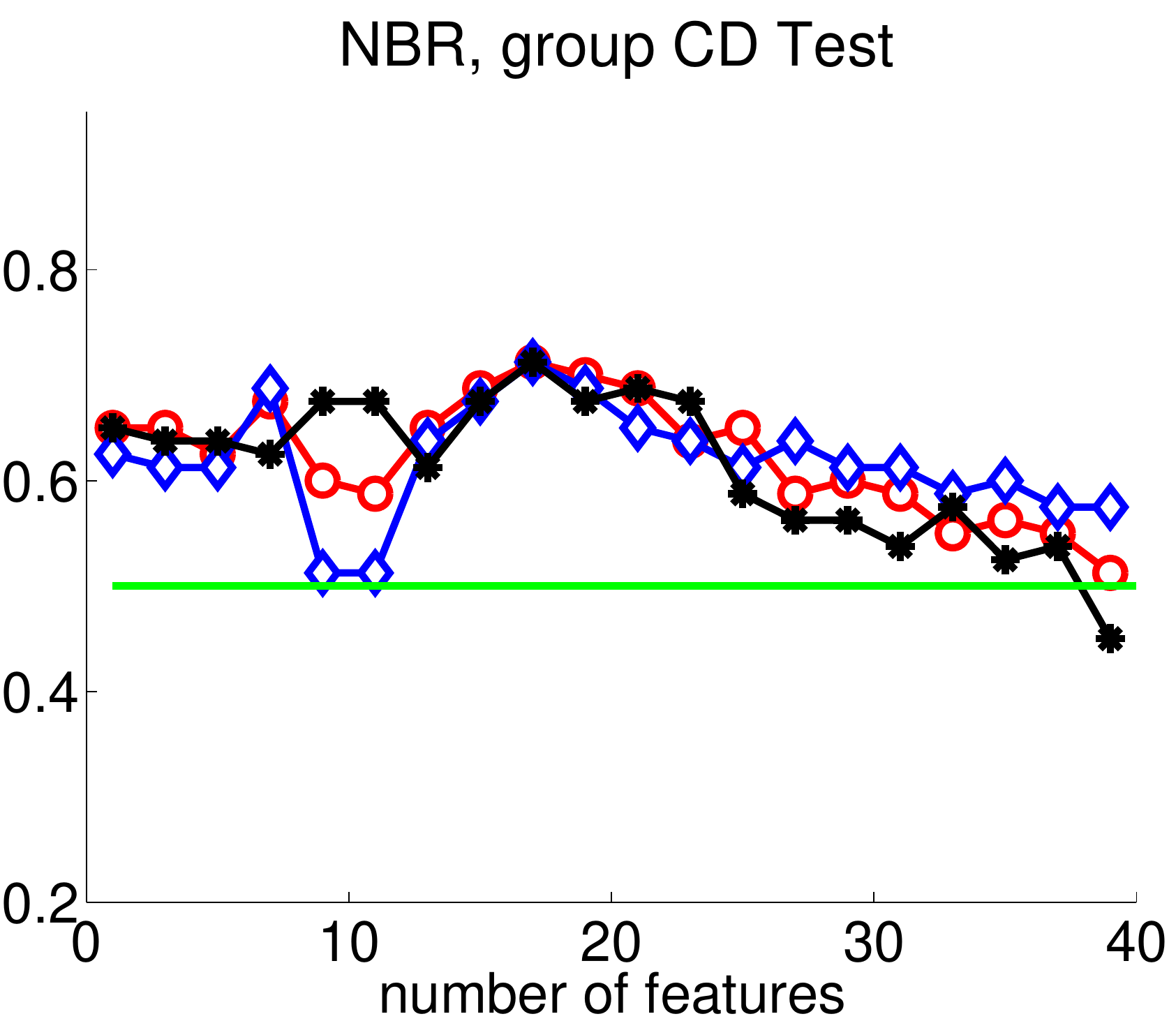}
\includegraphics[width=0.15\textwidth]{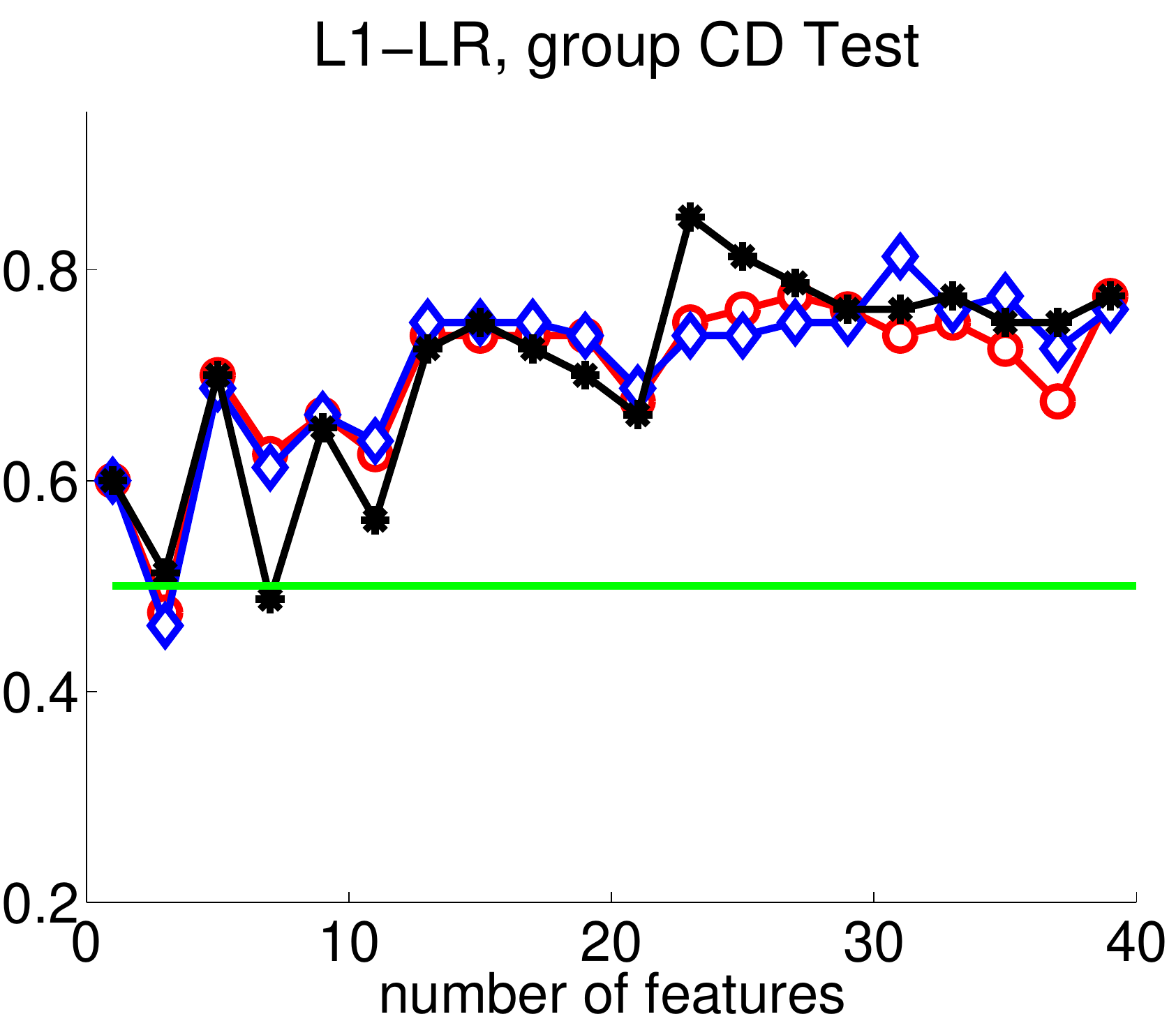}\\
\includegraphics[width=.4\textwidth]{images/legend.pdf}\\
\caption{Leave one experimental block out cross-validation accuracy for infant subjects as a function of the number of top ranked variables used for classification.  We use the same conventions of Fig. \ref{fig:adultEyeTrackResult}.}
\label{fig:infantEyeTrackResult}
\end{figure}

The three linear classifiers discussed above were applied to determine the LOBO-CV error as a function of the number of top features selected by the three different feature selection algorithms.  The results are shown in Fig.~\ref{fig:infantEyeTrackResult}, where we see that classifying infants requires significantly more variables than the adult case.  This is to be expected because of the diffuse looking pattern typical of babies.  The top five infant variables are shown in Table \ref{table:infantTesting}. The underlined entries were consistently selected by at least two feature selection algorithms and across both category conditions.  The consistent top variables in the learning and testing conditions were \emph{density of fixations} and \emph{DHB}, which describes the density of fixations at different distances from the relevant AOI(s). The \emph{fourth fixation} was also relevant in the testing condition.

%-----------------------------------------------
%-----------------------------------------------
%-----------------------------------------------
\begin{table}
%\centering
\scriptsize
% learning phase
\begin{tabular}{ r l }
\begin{sideways}\hspace{-.2cm}A or B\end{sideways} &\hspace{-.3cm} 
\begin{tabular}{ l  l  l }
\emph{ Learning condition} & & \\ \hline
\hline
\hline
~~~~ANOVA  & NBR &  L1-LR \\  \hline
1. \underline{Den of fix} at AOI 10 &  \underline{Den of fix} at AOI 2 &   \underline{Den of fix} at AOI 10  \\
2. $3^{rd}$ fix at AOI 10 & \underline{Den of fix} at AOI 10   & $1^{st}$ sac to AOI 5   \\ 
3. AOI 11, \underline{DHB} 5 & AOI 11, \underline{DHB} 5 &  \underline{Den of fix} at AOI 1    \\
4. Den of sac to AOI 10 & AOI 11, \underline{DHB} 35  &  $2^{nd}$ fix at AOI 1 \\ 
5. AOI 4, \underline{DHB} 20 & AOI 11, \underline{DHB} 22 &  \underline{Den of fix} at AOI 2  \\
6. $2^{nd}$ fix at AOI 10  &  Den of sac to AOI 2 &   AOI 11, \underline{DHB} 5  \\
7. $2^{nd}$ fix at AOI 1 &  \underline{Den of fix} at AOI 1   &   AOI 11, \underline{DHB} 22  \\ 
8.  \underline{Den of fix} at AOI 1  &  \underline{Den of fix} at AOI 9 &  $2^{nd}$ sac to AOI 3 \\
9.  \underline{Den of fix} at AOI 2 & AOI 4, \underline{DHB} 7 &   AOI 11, \underline{DHB} 16  \\ 
10.  AOI 11, \underline{DHB} 22 & AOI 4, \underline{DHB} 12  &  \underline{Den of fix} at AOI 3   \\
\hline 
\end{tabular}
\end{tabular}% end A or B tabular

\begin{tabular}{ r l }
\begin{sideways}\hspace{0cm}C or D\end{sideways} &\hspace{-.3cm} 
\begin{tabular}{l  l  l }
\hline
1. AOI 13, \underline{DHB} 5 & $4^{th}$ fix at AOI 2  &  AOI 13, \underline{DHB} 5    \\
2. AOI 6, \underline{DHB} 21 &  $3^{rd}$ sac to AOI 2  &   AOI 6, \underline{DHB} 21    \\ 
3. \underline{Den of fix} at AOI 13  & $1^{st}$ fix at AOI 5  &   \underline{Den of fix} at AOI 13  \\
4. AOI 13, \underline{DHB} 2 &  $3^{rd}$ sac to non-rel AOI &   $3^{rd}$ sac to AOI 10  \\ 
5. $3^{rd}$ sac to non-rel AOI  &  $3^{rd}$ fix at AOI 6  & $4^{th}$ fix at AOI 10   \\
6. $1^{st}$ fix at AOI 14 & $3^{rd}$ fix at AOI 9  &  $3^{rd}$ sac to non-rel AOI \\
7. $1^{st}$ fix at non-rel AOI & AOI 6, \underline{DHB} 21  &  $4^{th}$ fix at AOI 5 \\ 
8. \underline{Den of fix} at AOI 14  & AOI 13, \underline{DHB} 5 &   $1^{st}$ fix at AOI 14 \\
9. AOI 6, \underline{DHB} 8 &  $1^{st}$ fix at non-rel AOI &  $5^{th}$ fix at AOI 3  \\ 
10. $4^{th}$ fix at AOI 5 &  $4^{th}$ fix at AOI 4  &  $4^{th}$ sac to AOI 1 \\
\hline 
\hline
\hline
\end{tabular}
\end{tabular}%end c or D tabular

% testing phase
\begin{tabular}{ r l }
\begin{sideways}\hspace{-.2cm}A or B\end{sideways} &\hspace{-.3cm} 
\begin{tabular}{ l  l  l }
\emph{ Testing condition} & & \\ \hline
\hline
\hline
~~~~ANOVA & NBR &  L1-LR \\  \hline
1. $6^{th}$ fix at non-rel AOI  & $6^{th}$ fix at non-rel AOI  & $6^{th}$ fix at non-rel AOI   \\
2. AOI 4, \underline{DHB} 20  &  AOI 4, \underline{DHB} 20  & AOI 4, \underline{DHB} 8   \\ 
3. AOI 4, \underline{DHB} 8 &  \underline{Den of fix} at AOI 10   &    \underline{$4^{th}$ fix} at AOI 7 \\
4. \underline{$4^{th}$ fix} at AOI 7 & Den of sac to AOI 10  &  \underline{Den of fix} at AOI 13  \\ 
5. Den of sac to AOI 10 & \underline{$4^{th}$ fix} at AOI 7  & AOI 4, \underline{DHB} 20  \\
6. AOI 4, \underline{DHB} 10  &  AOI 4, \underline{DHB} 8 &  AOI 11, \underline{DHB} 16 \\
7. \underline{Den of fix} at AOI 10   &   AOI 4, \underline{DHB} 10  &  $7^{th}$ fix at AOI 7 \\ 
8. $7^{th}$ fix at AOI 10 &  number unique AOIs fixated   &  AOI 4, \underline{DHB} 10 \\
9. $1^{st}$ sac to AOI 10 & $1^{st}$ fix at AOI 13 & Den of sac to AOI 10 \\ 
10. $1^{st}$ fix at AOI 1 & $2^{nd}$ fix at AOI 14  &  \underline{Den of fix} at AOI 1   \\
\hline 
\end{tabular}
\end{tabular} %end A or B

\begin{tabular}{ r l }
\begin{sideways}\hspace{0cm}C or D\end{sideways} &\hspace{-.3cm} 
\begin{tabular}{ l  l  l }
\hline
1. \underline{Den of fix} at AOI 7 & \underline{Den of fix} at AOI 7   &  $3^{rd}$ sac to AOI 3  \\
2. $3^{rd}$ sac to AOI 3  &  $3^{rd}$ fix at AOI 7  & \underline{Den of fix} at AOI 7  \\ 
3.  $1^{st}$ fix at AOI 7 &  $6^{th}$ fix at AOI 7  &  \underline{$4^{th}$ fix} at AOI 10  \\
4.  \underline{$4^{th}$ fix} at AOI 10  &  Duration of $2^{nd}$ fix   &  \underline{$4^{th}$ fix} at AOI 12   \\ 
5.  $3^{rd}$ fix at AOI 7   & AOI 13, \underline{DHB} 2  &   $2^{nd}$ fix at AOI 1  \\
6. $6^{th}$ fix at AOI 7  & $3^{rd}$ sac to AOI 3 &  $2^{nd}$ sac to AOI 8 \\
7. $2^{nd}$ fix at AOI 1  &  $1^{st}$ fix at AOI 7  & $2^{nd}$ fix at AOI 2 \\ 
8. $2^{nd}$ fix at AOI 2 &  \underline{$4^{th}$ fix} at AOI 7  & $6^{th}$ fix at AOI 8 \\
9. $1^{st}$ sac to AOI 10  &  \underline{$4^{th}$ fix} at AOI 10  & $2^{nd}$ sac to AOI 1 \\ 
10. AOI 13, \underline{DHB} 12 & AOI 6, \underline{DHB} 7  & \underline{Den of fix} at AOI 1 \\
\hline 
\hline
\hline
\end{tabular}
\end{tabular} %end C or D
\caption{Infant Experiment: The following variables were determined most relevant during the category learning and category discrimination phases of the infant experiment.  The consistently selected variables are underlined.  We use the same conventions as Table \ref{table:adultTesting}.  }
\label{table:infantTesting}
\end{table}
%-----------------------------------------------

\subsection{Comparing Infants to Adults}\label{exp2}
The above results raise a new question. How similar are the attention models of adults and infants? Specifically, since the infant data are so noisy, can we use the adult model to improve on the infant one?  To test this, we used the adult SVM classifier model trained with the top five variables from ANOVA to predict if infants were learners or non-learners.  This was done only for the testing phase, because the testing phase images for adults and infants are similar so that the extracted variables correspond. Infants were classified with $49\%$ accuracy in the category A or B condition.  Infants were classified with $50\%$ accuracy in the category C or D condition.  These chance performance of the adult model identifying infant learners suggests that adults and infants attend to category objects differently. The remaining challenge is to examine the generality of this finding by testing a broader set of categories.
 
\section{Discussion}
The analysis demonstrates that the proposed method of variable selection is viable.  We can predict if adults have learned a category based on a very small number of top ranked eye track variables.  Furthermore, there is strong agreement between the different ranking approaches about which variables are most important.  Specifically, the consistently top ranked variables in the learning condition were \emph{latency to a fixation at the relevant AOI}, \emph{density of fixations at the relevant AOI}, and \emph{first fixation at the relevant AOI}. The consistently top ranked variables in the testing condition were \emph{first, second, and third fixations}.  These results suggest that during learning, adult category learners focus their attention on the relevant category features.  The results also suggest that adult category learners make discrimination judgments within the first few fixations.  

The infant data analysis also demonstrated that we can predict category learning, but it requires a larger number of variables.  Again, there was agreement between the different ranking approaches about which variables are most important.  The consistent top variables in the learning and testing conditions describe the \emph{fixation density} at different areas of the image. The \emph{fourth fixation} was also relevant in the testing condition.  These results suggest that for infants, the pattern of fixations over the entire object is more informative than the amount of time fixating the relevant AOI.  Therefore, it appears that whereas category learning in adults in marked by focused attention to category-relevant features, category learning in infants is marked by more diffused attention coupled with exploration of multiple areas of interest.
Finally, we showed that the adult model does not predict infant category learning.  We address these findings in the following sections.

\subsection{Why were the best variables different for infants and adults?} 
There is an important difference between the variable selection results of the adult experiment versus the infant experiment. Namely, while adult learners are identified readily with a small set of variables emphasizing early looks at the relevant AOI(s), infant learners are better identified based on their pattern of fixating over the trial.  We propose an explanation based on the goals of adult versus infant participants.

Although the experiment stimuli were the same for adults and infants, there were fundamental differences in the design of the experiments.  Namely, the objectives during the experiment were different for adults versus infants.  In the case of adults, the participants were given a particular task: learn how to identify a member of this category from a set of exemplars, then identify a member of that category from a pair of objects.  Therefore, the adults' goal was to learn the category object as quickly as possible given the limited number of training examples, such that discrimination could be performed accurately during the testing phase.  Given this goal, it was reasonable that the consistently selected variables were associated with relevant AOI fixation density as well as early looks (see Table \ref{table:adultTesting} ).  

In the case of infants, we used sound and motion to draw the infantÕs attention to the relevant AOI in hopes that he or she learned to identify the category object.  Then, we assumed that if the category was learned, the infants would show a preference for either the learned category or novel category during the discrimination phase. To this uncertainty, we ought to add the large amounts of random movements of the infant's gaze.  As we see in our results, a larger set of variables is required to reliably distinguishing learners from non-learners.  In addition, while fixation density is important, the emphasis is not on fixating the relevant AOI.  

\section{Conclusion}
We have developed a methodology for automatically determining eye tracking variables that are relevant to understanding category learning and discrimination processes.  Previous research has relied on ad-hoc techniques to determine which variables should be analyzed.  Instead, we used statistical methods to find the important variables in an over-complete set of variables.  

The efficacy of the approach was verified with an adult and infant categorization study.  The variables determined most relevant for adults emphasize looking at the relevant AOI(s) longer, and earlier during the categorization tasks.  This result is satisfying for two reasons: 1) It is expected that category learners quickly focus their efforts on the relevant AOI(s), and 2) these variables coincide with the variables \emph{proportion fixation time} and \emph{relative priority} of previous eye tracking category learning studies such as \cite{Rehder2005a}.  The variables determined most relevant for infants emphasize the overall pattern of fixating the object.  This result is also satisfying because infants are expected to explore objects. 

Note that the important variables were verified by the \emph{task} and \emph{stimuli} described. Altering these parameters may result in different important variables. By comparing the important variables among different tasks and stimuli, we can further dissociate which eye tracking variables are linked to specific processes during categorization. 

\section*{Acknowledgments}
This research was partially supported by NIH grant R01 EY-020834 to AM, NSF grant BCS-0720135 and NIH grant R01 HD-056105 to VS, and a Seed Grant by the Center for Cognitive Science (CCS) at OSU to DBW, VS, and AM. SR was partially supported by a fellowship from the CCS.

%-------------------------------------------------
% ,amso2006,johnson2008,Falk-Ytter:2006,Quinn:2009,McMurray:2004,best2010

% DONÂT TOUCH BELOW THIS
\nocite{Rayner1998}
\bibliography{bibFile}{}

\begin{thebibliography}{}

\bibitem [\protect \citeauthoryear {%
Amso%
\ \BBA {} Johnson%
}{%
Amso%
\ \BBA {} Johnson%
}{%
{\protect \APACyear {2005}}%
}]{%
amso2005}
\APACinsertmetastar {%
amso2005}%
\begin{APACrefauthors}%
Amso, D.%
\BCBT {}\ \BBA {} Johnson, S\BPBI P.%
\end{APACrefauthors}%
\unskip\
\newblock
\APACrefYearMonthDay{2005}{}{}.
\newblock
{\BBOQ}\APACrefatitle {Selection and inhibition in infancy: evidence from the
  spatial negative priming paradigm} {Selection and inhibition in infancy:
  evidence from the spatial negative priming paradigm}.{\BBCQ}
\newblock
\APACjournalVolNumPages{Cognition}{95}{2}{B27--B36}.
\PrintBackRefs{\CurrentBib}

\bibitem [\protect \citeauthoryear {%
Amso%
\ \BBA {} Johnson%
}{%
Amso%
\ \BBA {} Johnson%
}{%
{\protect \APACyear {2006}}%
}]{%
amso2006}
\APACinsertmetastar {%
amso2006}%
\begin{APACrefauthors}%
Amso, D.%
\BCBT {}\ \BBA {} Johnson, S\BPBI P.%
\end{APACrefauthors}%
\unskip\
\newblock
\APACrefYearMonthDay{2006}{}{}.
\newblock
{\BBOQ}\APACrefatitle {Learning by Selection: Visual Search and Object
  Perception in Young Infants} {Learning by selection: Visual search and object
  perception in young infants}.{\BBCQ}
\newblock
\APACjournalVolNumPages{Developmental Psychology}{42}{6}{1236--1245}.
\PrintBackRefs{\CurrentBib}

\bibitem [\protect \citeauthoryear {%
Amso%
\ \BBA {} Johnson%
}{%
Amso%
\ \BBA {} Johnson%
}{%
{\protect \APACyear {2008}}%
}]{%
amso2008}
\APACinsertmetastar {%
amso2008}%
\begin{APACrefauthors}%
Amso, D.%
\BCBT {}\ \BBA {} Johnson, S\BPBI P.%
\end{APACrefauthors}%
\unskip\
\newblock
\APACrefYearMonthDay{2008}{}{}.
\newblock
{\BBOQ}\APACrefatitle {Development of visual selection in 3- to 9-month-olds:
  Evidence from saccades to previously ignored locations} {Development of
  visual selection in 3- to 9-month-olds: Evidence from saccades to previously
  ignored locations}.{\BBCQ}
\newblock
\APACjournalVolNumPages{Infancy}{13}{}{675--686}.
\PrintBackRefs{\CurrentBib}

\bibitem [\protect \citeauthoryear {%
Best%
, Robinson%
\BCBL {}\ \BBA {} Sloutsky%
}{%
Best%
\ \protect \BOthers {.}}{%
{\protect \APACyear {2010}}%
}]{%
best2010}
\APACinsertmetastar {%
best2010}%
\begin{APACrefauthors}%
Best, C\BPBI A.%
, Robinson, C\BPBI W.%
\BCBL {}\ \BBA {} Sloutsky, V\BPBI M.%
\end{APACrefauthors}%
\unskip\
\newblock
\APACrefYearMonthDay{2010}{}{}.
\newblock
{\BBOQ}\APACrefatitle {The effect of labels on visual attention: An eye
  tracking study} {The effect of labels on visual attention: An eye tracking
  study}.{\BBCQ}
\newblock
\APACjournalVolNumPages{Proceedings of the 32nd Annual Conference of the
  Cognitive Science Society}{}{}{1846--1851}.
\PrintBackRefs{\CurrentBib}

\bibitem [\protect \citeauthoryear {%
Burges%
}{%
Burges%
}{%
{\protect \APACyear {1998}}%
}]{%
Burges1998}
\APACinsertmetastar {%
Burges1998}%
\begin{APACrefauthors}%
Burges, C\BPBI J\BPBI C.%
\end{APACrefauthors}%
\unskip\
\newblock
\APACrefYearMonthDay{1998}{}{}.
\newblock
{\BBOQ}\APACrefatitle {A Tutorial on Support Vector Machines for Pattern
  Recognition} {A tutorial on support vector machines for pattern
  recognition}.{\BBCQ}
\newblock
\APACjournalVolNumPages{Data Mining and Knowledge Discovery}{2}{}{121--167}.
\PrintBackRefs{\CurrentBib}

\bibitem [\protect \citeauthoryear {%
Chang%
\ \BBA {} Lin%
}{%
Chang%
\ \BBA {} Lin%
}{%
{\protect \APACyear {2001}}%
}]{%
LIBSVM}
\APACinsertmetastar {%
LIBSVM}%
\begin{APACrefauthors}%
Chang, C\BHBI C.%
\BCBT {}\ \BBA {} Lin, C\BHBI J.%
\end{APACrefauthors}%
\unskip\
\newblock
\APACrefYearMonthDay{2001}{}{}.
\newblock
{\BBOQ}\APACrefatitle {{LIBSVM}: a library for support vector machines}
  {{LIBSVM}: a library for support vector machines}{\BBCQ}\
  [\bibcomputersoftwaremanual].
\newblock
\APACrefnote{Software available at {http://www.csie.ntu.edu.tw/~cjlin/libsvm}}
\PrintBackRefs{\CurrentBib}

\bibitem [\protect \citeauthoryear {%
Duda%
, Hart%
\BCBL {}\ \BBA {} Stork%
}{%
Duda%
\ \protect \BOthers {.}}{%
{\protect \APACyear {2001}}%
}]{%
duda2001}
\APACinsertmetastar {%
duda2001}%
\begin{APACrefauthors}%
Duda, R\BPBI O.%
, Hart, P\BPBI E.%
\BCBL {}\ \BBA {} Stork, D\BPBI G.%
\end{APACrefauthors}%
\unskip\
\newblock
\APACrefYear{2001}.
\newblock
\APACrefbtitle {Pattern Classification (2nd Edition)} {Pattern classification
  (2nd edition)}\ (\PrintOrdinal{2}\ \BEd).
\newblock
\APACaddressPublisher{}{Wiley-Interscience}.
\newblock
\APAChowpublished {Hardcover}.
\PrintBackRefs{\CurrentBib}

\bibitem [\protect \citeauthoryear {%
Falck-Ytter%
, Gredeb\"{a}ck%
\BCBL {}\ \BBA {} von Hofsten%
}{%
Falck-Ytter%
\ \protect \BOthers {.}}{%
{\protect \APACyear {2006}}%
}]{%
Falk-Ytter:2006}
\APACinsertmetastar {%
Falk-Ytter:2006}%
\begin{APACrefauthors}%
Falck-Ytter, T.%
, Gredeb\"{a}ck, G.%
\BCBL {}\ \BBA {} von Hofsten, C.%
\end{APACrefauthors}%
\unskip\
\newblock
\APACrefYearMonthDay{2006}{}{}.
\newblock
{\BBOQ}\APACrefatitle {Infants Predict Other People's Action Goals} {Infants
  predict other people's action goals}{\BBCQ}\ [Journal article].
\newblock
\APACjournalVolNumPages{Nature Neuroscience}{9}{7}{878--879}.
\PrintBackRefs{\CurrentBib}

\bibitem [\protect \citeauthoryear {%
Johnson%
, Amso%
\BCBL {}\ \BBA {} Slemmer%
}{%
Johnson%
\ \protect \BOthers {.}}{%
{\protect \APACyear {2003}}%
}]{%
johnson2003}
\APACinsertmetastar {%
johnson2003}%
\begin{APACrefauthors}%
Johnson, S\BPBI P.%
, Amso, D.%
\BCBL {}\ \BBA {} Slemmer, J\BPBI A.%
\end{APACrefauthors}%
\unskip\
\newblock
\APACrefYearMonthDay{2003}{}{}.
\newblock
{\BBOQ}\APACrefatitle {{Development of object concepts in infancy: Evidence for
  early learning in an eye-tracking paradigm}} {{Development of object concepts
  in infancy: Evidence for early learning in an eye-tracking paradigm}}.{\BBCQ}
\newblock
\APACjournalVolNumPages{Proceedings of the National Academy of
  Sciences}{100}{18}{10568--10573}.
\PrintBackRefs{\CurrentBib}

\bibitem [\protect \citeauthoryear {%
Johnson%
, Davidow%
, Hall-Haro%
\BCBL {}\ \BBA {} Frank%
}{%
Johnson%
\ \protect \BOthers {.}}{%
{\protect \APACyear {2008}}%
}]{%
johnson2008}
\APACinsertmetastar {%
johnson2008}%
\begin{APACrefauthors}%
Johnson, S\BPBI P.%
, Davidow, J.%
, Hall-Haro, C.%
\BCBL {}\ \BBA {} Frank, M\BPBI C.%
\end{APACrefauthors}%
\unskip\
\newblock
\APACrefYearMonthDay{2008}{}{}.
\newblock
{\BBOQ}\APACrefatitle {Development of perceptual completion originates in
  information acquisition} {Development of perceptual completion originates in
  information acquisition}.{\BBCQ}
\newblock
\APACjournalVolNumPages{Developmental Psychology}{44}{5}{1214--1224}.
\PrintBackRefs{\CurrentBib}

\bibitem [\protect \citeauthoryear {%
Johnson%
, Slemmer%
\BCBL {}\ \BBA {} Amso%
}{%
Johnson%
\ \protect \BOthers {.}}{%
{\protect \APACyear {2004}}%
}]{%
johnson2004}
\APACinsertmetastar {%
johnson2004}%
\begin{APACrefauthors}%
Johnson, S\BPBI P.%
, Slemmer, J\BPBI A.%
\BCBL {}\ \BBA {} Amso, D.%
\end{APACrefauthors}%
\unskip\
\newblock
\APACrefYearMonthDay{2004}{}{}.
\newblock
{\BBOQ}\APACrefatitle {Where infants look determines how they see: Eye
  movements and object perception performance in 3-month-olds} {Where infants
  look determines how they see: Eye movements and object perception performance
  in 3-month-olds}.{\BBCQ}
\newblock
\APACjournalVolNumPages{Infancy}{6}{2}{185--201}.
\PrintBackRefs{\CurrentBib}

\bibitem [\protect \citeauthoryear {%
Kloos%
\ \BBA {} Sloutsky%
}{%
Kloos%
\ \BBA {} Sloutsky%
}{%
{\protect \APACyear {2008}}%
}]{%
Kloos2008}
\APACinsertmetastar {%
Kloos2008}%
\begin{APACrefauthors}%
Kloos, H.%
\BCBT {}\ \BBA {} Sloutsky, V\BPBI M.%
\end{APACrefauthors}%
\unskip\
\newblock
\APACrefYearMonthDay{2008}{}{}.
\newblock
{\BBOQ}\APACrefatitle {What's behind different kinds of kinds: Effects of
  statistical density on learning and representation of categories} {What's
  behind different kinds of kinds: Effects of statistical density on learning
  and representation of categories}.{\BBCQ}
\newblock
\APACjournalVolNumPages{Journal of Experimental Psychology:
  General}{137}{1}{52-72}.
\PrintBackRefs{\CurrentBib}

\bibitem [\protect \citeauthoryear {%
Martinez%
\ \BBA {} Zhu%
}{%
Martinez%
\ \BBA {} Zhu%
}{%
{\protect \APACyear {2005}}%
}]{%
martinez2005}
\APACinsertmetastar {%
martinez2005}%
\begin{APACrefauthors}%
Martinez, A\BPBI M.%
\BCBT {}\ \BBA {} Zhu, M.%
\end{APACrefauthors}%
\unskip\
\newblock
\APACrefYearMonthDay{2005}{}{}.
\newblock
{\BBOQ}\APACrefatitle {Where are linear feature extraction methods applicable?}
  {Where are linear feature extraction methods applicable?}{\BBCQ}
\newblock
\APACjournalVolNumPages{Pattern Analysis and Machine Intelligence, IEEE
  Transactions on}{27}{12}{1934--1944}.
\PrintBackRefs{\CurrentBib}

\bibitem [\protect \citeauthoryear {%
McMurray%
\ \BBA {} Aslin%
}{%
McMurray%
\ \BBA {} Aslin%
}{%
{\protect \APACyear {2004}}%
}]{%
McMurray:2004}
\APACinsertmetastar {%
McMurray:2004}%
\begin{APACrefauthors}%
McMurray, B.%
\BCBT {}\ \BBA {} Aslin, R\BPBI N.%
\end{APACrefauthors}%
\unskip\
\newblock
\APACrefYearMonthDay{2004}{}{}.
\newblock
{\BBOQ}\APACrefatitle {Anticipatory Eye Movements Reveal Infants' Auditory and
  Visual Categories} {Anticipatory eye movements reveal infants' auditory and
  visual categories}.{\BBCQ}
\newblock
\APACjournalVolNumPages{Infancy}{6}{2}{203--229}.
\PrintBackRefs{\CurrentBib}

\bibitem [\protect \citeauthoryear {%
Murphy%
}{%
Murphy%
}{%
{\protect \APACyear {2004}}%
}]{%
kalmanToolbox}
\APACinsertmetastar {%
kalmanToolbox}%
\begin{APACrefauthors}%
Murphy, K.%
\end{APACrefauthors}%
\unskip\
\newblock
\APACrefYearMonthDay{2004}{}{}.
\newblock
\APACrefbtitle {Kalman filter toolbox for Matlab.} {Kalman filter toolbox for
  matlab.}
\newblock
\begin{APACrefURL}
  \url{http://www.cs.ubc.ca/~murphyk/Software/Kalman/kalman.html}
  \end{APACrefURL}
\PrintBackRefs{\CurrentBib}

\bibitem [\protect \citeauthoryear {%
Quinn%
, Doran%
, Reiss%
\BCBL {}\ \BBA {} Hoffman%
}{%
Quinn%
\ \protect \BOthers {.}}{%
{\protect \APACyear {2009}}%
}]{%
Quinn:2009}
\APACinsertmetastar {%
Quinn:2009}%
\begin{APACrefauthors}%
Quinn, P\BPBI C.%
, Doran, M\BPBI M.%
, Reiss, J\BPBI E.%
\BCBL {}\ \BBA {} Hoffman, J\BPBI E.%
\end{APACrefauthors}%
\unskip\
\newblock
\APACrefYearMonthDay{2009}{}{}.
\newblock
{\BBOQ}\APACrefatitle {Time course of visual attention in infant categorization
  of cats versus dogs: evidence for a head bias as revealed through eye
  tracking.} {Time course of visual attention in infant categorization of cats
  versus dogs: evidence for a head bias as revealed through eye
  tracking.}{\BBCQ}
\newblock
\APACjournalVolNumPages{Child development}{80}{1}{151--161}.
\PrintBackRefs{\CurrentBib}

\bibitem [\protect \citeauthoryear {%
Quinn%
, Eimas%
\BCBL {}\ \BBA {} Rosenkrantz%
}{%
Quinn%
\ \protect \BOthers {.}}{%
{\protect \APACyear {1993}}%
}]{%
Quinn1993}
\APACinsertmetastar {%
Quinn1993}%
\begin{APACrefauthors}%
Quinn, P\BPBI C.%
, Eimas, P\BPBI D.%
\BCBL {}\ \BBA {} Rosenkrantz, S\BPBI L.%
\end{APACrefauthors}%
\unskip\
\newblock
\APACrefYearMonthDay{1993}{}{}.
\newblock
{\BBOQ}\APACrefatitle {Evidence for representations of perceptually similar
  natural categories by 3-month-old and 4-month-old infants.} {Evidence for
  representations of perceptually similar natural categories by 3-month-old and
  4-month-old infants.}{\BBCQ}
\newblock
\APACjournalVolNumPages{Perception}{22}{4}{463--475}.
\PrintBackRefs{\CurrentBib}

\bibitem [\protect \citeauthoryear {%
Rayner%
}{%
Rayner%
}{%
{\protect \APACyear {1998}}%
}]{%
Rayner1998}
\APACinsertmetastar {%
Rayner1998}%
\begin{APACrefauthors}%
Rayner, K.%
\end{APACrefauthors}%
\unskip\
\newblock
\APACrefYearMonthDay{1998}{}{}.
\newblock
{\BBOQ}\APACrefatitle {Eye movements in reading and information processing: 20
  years of research.} {Eye movements in reading and information processing: 20
  years of research.}{\BBCQ}
\newblock
\APACjournalVolNumPages{Psychological Bulletin}{124}{3}{372--422}.
\PrintBackRefs{\CurrentBib}

\bibitem [\protect \citeauthoryear {%
Rehder%
\ \BBA {} Hoffman%
}{%
Rehder%
\ \BBA {} Hoffman%
}{%
{\protect \APACyear {2005}}%
}]{%
Rehder2005a}
\APACinsertmetastar {%
Rehder2005a}%
\begin{APACrefauthors}%
Rehder, B.%
\BCBT {}\ \BBA {} Hoffman, A\BPBI B.%
\end{APACrefauthors}%
\unskip\
\newblock
\APACrefYearMonthDay{2005}{}{}.
\newblock
{\BBOQ}\APACrefatitle {Eyetracking and selective attention in category
  learning} {Eyetracking and selective attention in category learning}.{\BBCQ}
\newblock
\APACjournalVolNumPages{Cognitive Psychology}{51}{}{1-41}.
\PrintBackRefs{\CurrentBib}

\bibitem [\protect \citeauthoryear {%
Salvucci%
\ \BBA {} Goldberg%
}{%
Salvucci%
\ \BBA {} Goldberg%
}{%
{\protect \APACyear {2000}}%
}]{%
Salvucci2000}
\APACinsertmetastar {%
Salvucci2000}%
\begin{APACrefauthors}%
Salvucci, D\BPBI D.%
\BCBT {}\ \BBA {} Goldberg, J\BPBI H.%
\end{APACrefauthors}%
\unskip\
\newblock
\APACrefYearMonthDay{2000}{}{}.
\newblock
{\BBOQ}\APACrefatitle {Identifying fixations and saccades in eye-tracking
  protocols} {Identifying fixations and saccades in eye-tracking
  protocols}.{\BBCQ}
\newblock
\BIn{} \APACrefbtitle {ETRA '00: Proceedings of the 2000 symposium on Eye
  tracking research \& applications} {Etra '00: Proceedings of the 2000
  symposium on eye tracking research \& applications}\ (\BPGS\ 71--78).
\newblock
\APACaddressPublisher{New York, NY, USA}{}.
\PrintBackRefs{\CurrentBib}

\bibitem [\protect \citeauthoryear {%
Schmidt%
}{%
Schmidt%
}{%
{\protect \APACyear {2011}}%
}]{%
L1toolbox}
\APACinsertmetastar {%
L1toolbox}%
\begin{APACrefauthors}%
Schmidt, M.%
\end{APACrefauthors}%
\unskip\
\newblock
\APACrefYearMonthDay{2011}{}{}.
\newblock
\APACrefbtitle {L1General - Matlab code for solving L1-regularization
  problems.} {L1general - matlab code for solving l1-regularization problems.}
\newblock
\begin{APACrefURL} \url{http://www.di.ens.fr/~mschmidt/Software/L1General.zip}
  \end{APACrefURL}
\PrintBackRefs{\CurrentBib}

\bibitem [\protect \citeauthoryear {%
Stampe%
}{%
Stampe%
}{%
{\protect \APACyear {1993}}%
}]{%
stampe1993}
\APACinsertmetastar {%
stampe1993}%
\begin{APACrefauthors}%
Stampe, D\BPBI M.%
\end{APACrefauthors}%
\unskip\
\newblock
\APACrefYearMonthDay{1993}{}{}.
\newblock
{\BBOQ}\APACrefatitle {Heuristic filtering and reliable calibration methods for
  video-based pupil-tracking systems} {Heuristic filtering and reliable
  calibration methods for video-based pupil-tracking systems}.{\BBCQ}
\newblock
\APACjournalVolNumPages{Behavioral Research Methods, Instruments, \&
  Computers}{25}{2}{137--142}.
\PrintBackRefs{\CurrentBib}

\end{thebibliography}

\end{document}